\documentclass[aps,twocolumn,superscriptaddress,pra,amsmath,showkeys]{revtex4-1}
\usepackage[colorlinks=true,linkcolor=blue,urlcolor=blue,citecolor=blue]{hyperref}
\usepackage{graphicx}
\usepackage{cleveref}

\makeatletter
\renewcommand{\fnum@figure}{Figure \thefigure}
\makeatother

\begin{document}

\title{Engineering Nanoparticles with Pure High-Order Multipole Scattering}

\author{Vladimir A. Zenin}
\email{zenin@mci.sdu.dk}
\affiliation{Centre for Nano Optics, University of Southern Denmark, 5230 Odense, Denmark}
\affiliation{Contributed equally to this work}

\author{Cesar E. Garcia-Ortiz}
\affiliation{CICESE, Unidad Monterrey, Alianza Centro 504, PIIT Apodaca, NL 66629, Mexico}
\affiliation{Contributed equally to this work}

\author{Andrey B. Evlyukhin}
\email{a.b.evlyukhin@daad-alumni.de}
\affiliation{Institute of Quantum Optics, Leibniz University Hannover, 30167 Hannover, Germany}

\author{Yuanqing Yang}
\affiliation{Centre for Nano Optics, University of Southern Denmark, 5230 Odense, Denmark}

\author{Radu Malureanu}
\affiliation{Department of Photonics Engineering, Technical University of Denmark, 2800 Kgs. Lyngby, Denmark}
\affiliation{National Centre for Micro- and Nano-Fabrication, Technical University of Denmark, 2800 Kgs. Lyngby, Denmark}

\author{Sergey M. Novikov}
\affiliation{Centre for Nano Optics, University of Southern Denmark, 5230 Odense, Denmark}
\affiliation{Center for Photonics and 2D Materials, Moscow Institute of Physics and Technology, 141700 Dolgoprudny, Russia}

\author{Victor Coello}
\affiliation{CICESE, Unidad Monterrey, Alianza Centro 504, PIIT Apodaca, NL 66629, Mexico}

\author{Boris N. Chichkov}
\affiliation{Institute of Quantum Optics, Leibniz University Hannover, 30167 Hannover, Germany}
\affiliation{Lebedev Physical Institute, 119333 Moscow, Russia}

\author{Sergey I. Bozhevolnyi}
\affiliation{Centre for Nano Optics, University of Southern Denmark, 5230 Odense, Denmark}
\affiliation{Danish Institute for Advanced Study, University of Southern Denmark,  5230 Odense, Denmark}

\author{Andrei V. Lavrinenko}
\affiliation{Department of Photonics Engineering, Technical University of Denmark, 2800 Kgs. Lyngby, Denmark}

\author{N. Asger Mortensen}
\affiliation{Centre for Nano Optics, University of Southern Denmark, 5230 Odense, Denmark}
\affiliation{Danish Institute for Advanced Study, University of Southern Denmark, 5230 Odense, Denmark}

\date{\today}

\keywords{multipole decomposition, all-dielectric nanoparticles, scattering diagram, octopole, hexadecapole}

\begin{abstract}
The ability to control scattering directionality of nanoparticles is in high demand for many nanophotonic applications. One of the challenges is to design nanoparticles producing pure high-order multipole scattering (e.g., octopole, hexadecapole), whose contribution is usually negligible compared to strong low-order multipole scattering (i.e., dipole or quadrupole). Here we present an intuitive way to design such nanoparticles by introducing a void inside them. We show that both shell and ring nanostructures allow regimes with nearly pure high-order multipole scattering. Experimentally measured scattering diagrams from properly designed silicon rings at near-infrared wavelengths ($\sim$800 nm) reproduce well scattering patterns of an electric octopole and magnetic hexadecapole. Our findings advance significantly inverse engineering of nanoparticles from given complex scattering characteristics, with possible applications in biosensing, optical metasurfaces, and quantum communications.

(This document is the unedited Author's version of a work submitted to \textit{ACS Photonics} on January 16th, 2020. Feel free to criticize ;)
\end{abstract}

\maketitle

\begin{figure}
\centering\includegraphics{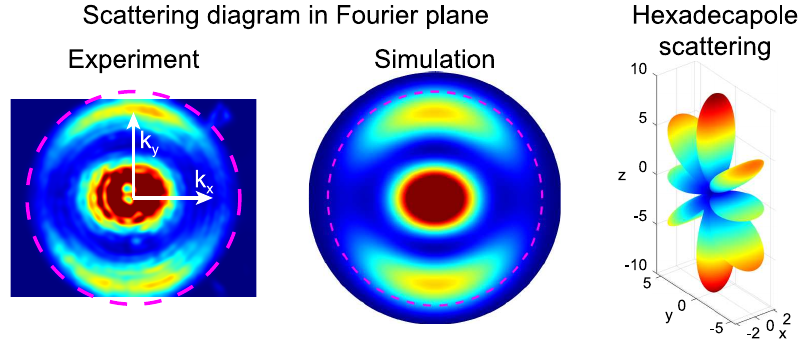}
\end{figure}

Modern photonic applications involve manipulation of light at the nanoscale by means of optical resonances. There are two main families of such resonances: polaritonic resonances originating from strongly-dispersive  negative dielectric permittivities (for example, plasmon-polariton resonances, supported by metallic nanoparticles), and photonic resonances utilizing high-refractive-index dielectric materials \cite{Evlyukhin2010, NL2012}. The latter is highly beneficial due to low, almost negligible absorption losses, compatibility with well-established semiconductor fabrication processes, and abundance of different optical modes (and corresponding resonances) even for simple symmetric shapes of dielectric nanoparticles \cite{diel1,diel2,fabr}. The above advantages led to a broad variety of applications utilizing dielectric nanoparticles, including light manipulation with metasurfaces \cite{meta1,meta2,meta3,meta4,meta5,meta6, metaR}, color printing \cite{color1,color2}, lasing \cite{lasers1,lasers2}, biosensing \cite{bio1,bio2,bio3}, strong coupling \cite{Strong_coupling1,Strong_coupling2,Strong_coupling3,Strong_coupling4}, and applications within quantum optics and topological photonics \cite{quant1,quant2,quant3}.

The optical properties of nanoparticles can be analyzed in different ways. One of the analytical tools is the multipolar decomposition, in which a generally complex field scattered by a nanoantenna is replaced by the superposition of fields (with relatively simple patterns) generated by basic point sources, called multipole moments, corresponding to the nanoantenna`s current distributions \cite{Jackson}. The number and type of multipole moments, which are sufficient to faithfully describe the scattered fields, are determined by the size, shape, and composition of nanoantennas. There are two basic approaches to the multipole decomposition of the scattered fields. The first is obtained from the Taylor expansion for the retarded potentials of electromagnetic fields generated by the induced electric currents in the nanoantennas. In this case, the multipole moments are determined as coefficients of the expansion and include ordinary multipole moments \cite{Raab} and the so-called mean-square radii \cite{Rodescu, Dubovik, Basharin, Beccherelli} or high-order toroidal moments \cite{Gurvitz2019}. The far-field nanoantenna scattering is, on the other hand, rational to describe in terms of angular distributions in the spherical coordinate system with the nanoantenna being in its center. Therefore, the second approach is based on the decomposition of a scattered far field into a series of the spherical harmonics, which form a natural basis in the spherical coordinate system and are assigned to the corresponding spherical multipoles \cite{Jackson}. In this case, the multipole moments are directly calculated from the distribution of scattered electric field on any spherical surface enclosing the nanoantenna \cite{Grahn}. By expressing the generated field on the spherical surface through the source currents, the spherical multipole moments can also be calculated using the current distributions induced inside the nanoantenna \cite{Jackson,Grahn}.

Different spherical multipoles generate the electromagnetic fields with different angular far-field distributions corresponding to given combinations of the spherical harmonics. In contrast to the spherical multipole presentation, the Taylor expansion approach provides sets of multipole moments (sometimes called Cartesian to differentiate from spherical multipoles), which might generate electromagnetic fields with \textit{identical} angular far-field distributions. For example, electric dipole, toroidal dipole and high-order toroidal dipole moments generate the electromagnetic fields with the identical angular far-field distribution corresponding to that generated by the spherical dipole moment \cite{Evlyukhin2016, Gurvitz2019}. Simply speaking, spherical multipole moments can be calculated by properly summing all Cartesian multipole moments having the same radiation pattern, while the reverse calculation is impossible. Recently however, explicit expressions for the spherical multipole moments have been found, from which the Cartesian multipole moments can be obtained by a simple Taylor expansion of these expressions \cite{Rockschtul2018, Rockschtul2019}. Finally, it was shown that the multipole family (Cartesian or spherical) is defined by the way how the expression for the scattered field is expanded: Taylor expansion leads to the Cartesian multipole moments, and expansion into spherical harmonics leads to the spherical multipole moments \cite{Evlyukhin2019}. In this work, we focus only on the far-field scattering, therefore only the spherical multipole decomposition is applied.

Similarly to Taylor series, where the first couple of terms represents usually the largest contribution in the expansion, only low-order multipole terms (dipoles, quadrupoles) contribute most to the total scattering from small nanoantennas, while contributions from high-order multipole terms (octopoles, hexadecapoles, and so on) are generally negligible. It is thus a non-trivial problem to find a nanoparticle, whose scattering is dominated by a contribution from high-order multipole moments. One way to get a considerable high-order multipole response is to increase the scatterer size. However, even for large scatterers, the high-order multipole contribution appears to be weak compared to strong contributions from the low-order multipole moments. Here, we solve the problem by introducing a void inside a nanoparticle without modifying its external dimensions. First, using numerical simulations we consider the evolution of multipole moments when increasing the diameter of a concentric spherical void inside a dielectric sphere, transforming eventually the latter into a shell nanoparticle. We find that, by introducing the void, the total scattering strength as well as the contribution from each multipole moment decreases. However, the reduction is stronger for the low-order multipole terms so that the relative contribution from the high-order multipole moments grows with the increase of the void. The same trend is also found for disk nanostructures, and we show that certain Si ring nanostructures scatter practically as the pure electric octopole or magnetic hexadecapole at the wavelength of 800 nm. This is verified by direct experimental measurements of the scattering diagram from individual nanoparticles. Our findings can be applied in biosensing, where narrow-band resonances and complex scattering patterns can boost the sensitivity and resolution. In a more general sense, our results significantly advance inverse engineering of nanoparticles, where the nanoparticle shape is to be obtained from given scattering properties. Finally, nanoparticles with such peculiar scattering properties can advantageously be used as meta-atoms for the design of metasurfaces exhibiting required complex functionalities. Particularly, it was recently shown that the high-order multipole term of scatterers is required in order to achieve full $2\pi$ phase coverage of Huygens' metasurface elements \cite{metaR}.

\section*{Scattering of sphere and shell nanoparticles}

\begin{figure*}[!htb]
\centering\includegraphics{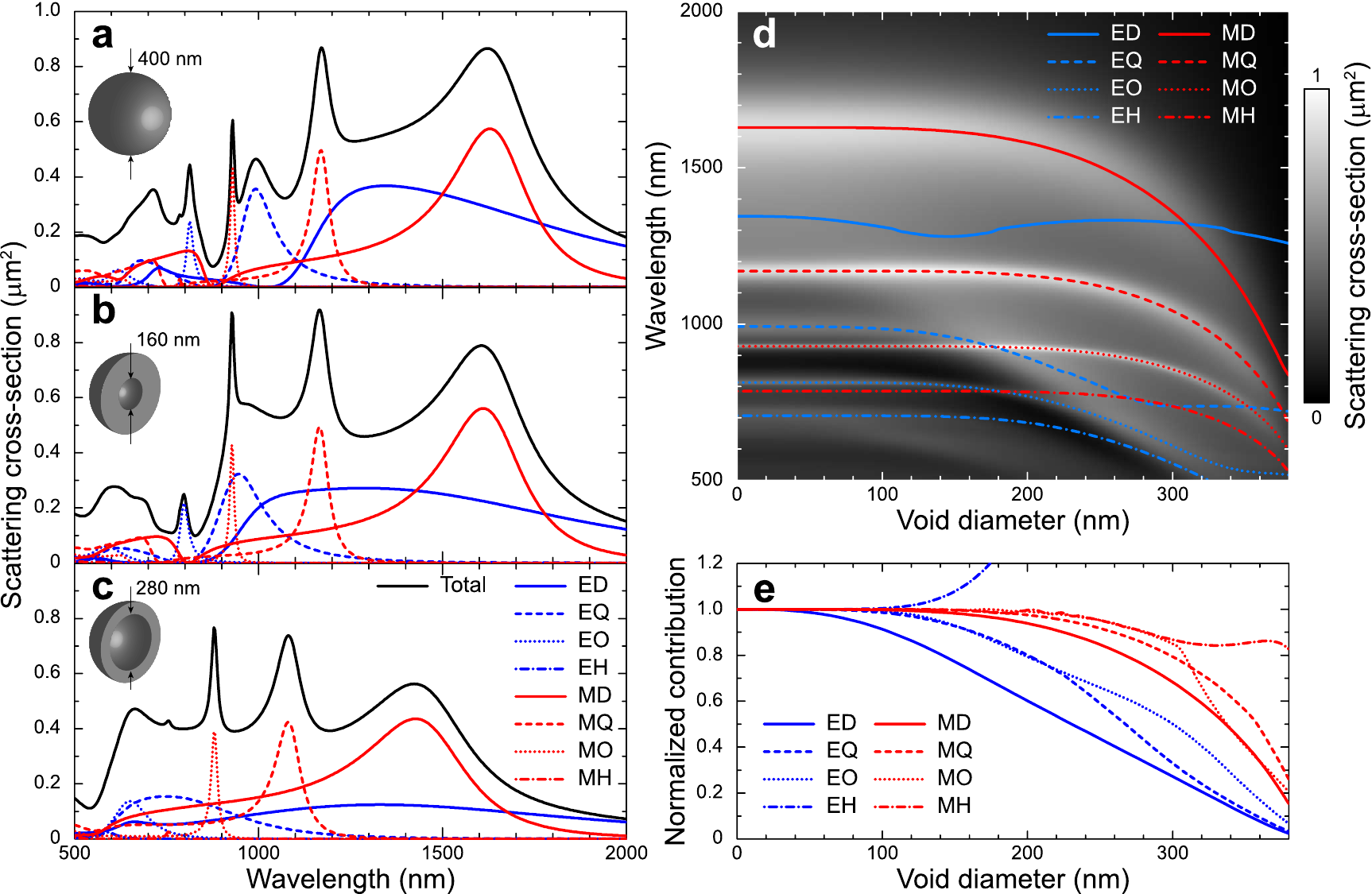}
  \caption{Evolution of scattering upon nanoparticle transformation from a solid sphere into a shell. (a-c) Total scattering cross-section (black) and individual contributions from electric (blue) and magnetic multipoles (red) for a silicon sphere without a void (a) and with a void diameter of 160 nm (b) and 280 nm (c). Here ED, EQ, EO, and EH stand for electrical dipole, quadrupole, octopole and hexadecapole, respectively, while MD, MQ, MO, MH stand for the corresponding magnetic multipoles. The outer diameter of Si sphere/shell is 400 nm. The refractive index of the surroundings and inside the void is assumed to be $n = 1.48$. (d) Total SCS as a function of the free-space wavelength and void diameter for the Si sphere/shell particle. Resonances for electric (blue) and magnetic multipoles (red) are shown with lines. (e) Normalized contributions of electric (blue) and magnetic multipoles (red) to the total SCS along their resonance curves as a function of the void diameter. Normalization is done to their contribution at zero void size (i.e., for the solid sphere).}
  \label{fig1}
\end{figure*}

We begin by numerically analyzing the scattering produced by a Si nanoparticle of the simplest morphology -- a sphere, where we introduce a void in the center and gradually transform the solid sphere into a shell (Figure~\ref{fig1}). Throughout the rest of the work, the refractive index of Si is taken from measurements of a deposited amorphous Si (see Supporting information, Figure~\ref{S1}), and the outer diameter of the sphere and shell particles is fixed to 400 nm. Mie theory \cite{Mie,Bohren} was used to calculate the total scattering cross-section (SCS) and multipole decomposition (see Supporting information, Figures~\ref{S2}-\ref{S3}). We found that the contribution from each multipole (except for the electric dipole) to the total SCS has a resonance-shape dependence on the wavelength, i.e., there is a well-defined peak (see Supporting information, Figure~\ref{S2}), whose position was determined for every void diameter and plotted on top of the total SCS map (Figure~\ref{fig1}d). These resonances are narrow and well separated, so one can get nearly pure high-order multipole scattering (electric and magnetic quadrupole, and magnetic octopole) for a solid Si sphere (Figure~\ref{fig1}a). A nearly pure electric octopole (EO) scattering can be found for the Si shell particle with the void diameter of 160 nm (Figure~\ref{fig1}b, at $\lambda \approx 800$ nm). Finally, by increasing the void diameter further to 280 nm, one can enable a magnetic octopole (MO) resonance well-separated from all other multipole resonances, thus having its dominant contribution in a broad wavelength range (Figure~\ref{fig1}c, at $\lambda \approx 870$ nm). The dominant multipole and its relative contribution to the total SCS for varied void diameter can be found in Supporting information, Figure~\ref{S3}.

In order to analyze the influence of the void size on the response of each multipole, we calculated their contribution to the total SCS at the resonance wavelength of the multipole, and normalized to its contribution at zero void size (Figure~\ref{fig1}e). By increasing the void size, the total SCS and contributions from each multipole at its resonance decrease with a blueshift of their peaks (Figure~\ref{fig1}d,e). This is expected, since the total volume of Si is reduced. However, the scattering contributions from the low-order multipoles decrease faster compared to the high-order multipoles. Additionally, the contributions from electric multipoles decrease faster compared to magnetic multipoles. This can be explained as following: the multipole moment is proportional to the volume integral of $j_m \left( kr \right) \Pi  \left( \mathbf{r}^m,\mathbf{j} \right) / r^m$, where $j_m$ is the $m$-th order spherical Bessel function of the first kind, $k = 2\pi n/\lambda$ is the wavenumber in the surrounding environment, $\mathbf{j}$ is the induced electric current density, $\mathbf{r}$ is the radius vector of the observation point, $\Pi  \left( \mathbf{r}^m,\mathbf{j} \right)$ is the combination of vector and scalar product between $\mathbf{j}$ and $m$-times $\mathbf{r}$, and the integration is done over the volume of the particle \cite{Rockschtul2018,Evlyukhin2019}. In the above expression $m = n - 1$ for electric and $m = n$ for magnetic multipoles of order $n$ ($n=1$ means dipole, $n=2$ - quadrupole, etc.). Thus, the spherical Bessel function acts as a weight inside the integration, therefore the lower the multipole order, the smaller it is influenced by the suppression of the central part. The zero-order spherical Bessel function is the only one having non-zero value at the origin (see Supporting information, Figure~\ref{S2}i), therefore it is only the electric dipole contribution, which is significantly influenced by the introduction of a small void (Figure~\ref{fig1}e).

\begin{figure*}[!htb]
\centering\includegraphics{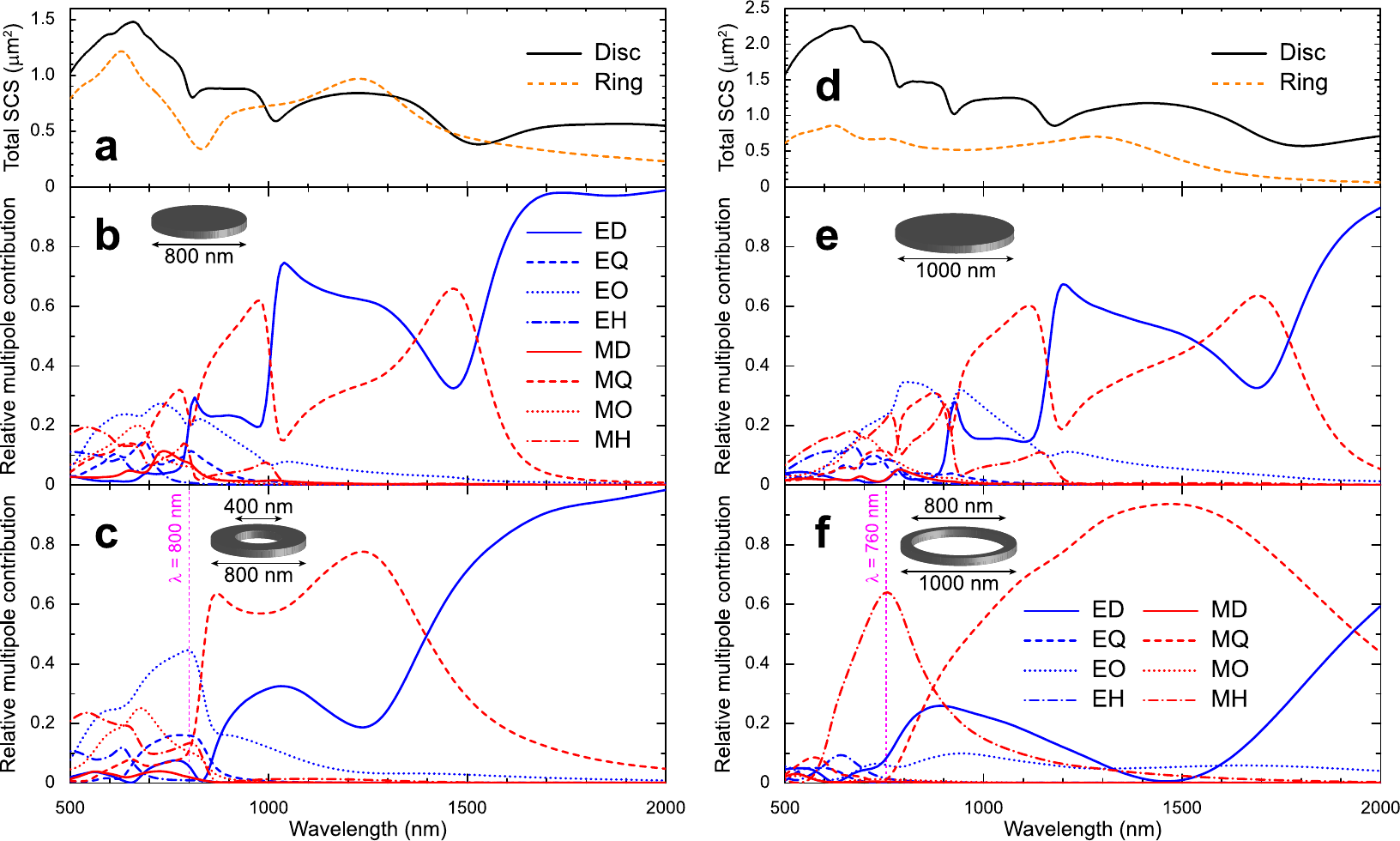}
  \caption{
Multipole analysis of disk and ring scattering. (a,d) Total SCS of the disk (solid black) and ring (dashed orange) with (a) 400/800 nm and (d) 800/1000 nm inner/outer diameter, respectively. (b,c) Relative contributions from electric (blue) and magnetic multipoles (red) to the total SCS for the disk (b) and ring (c) with the inner/outer diameter of 400/800 nm. (e,f) Relative contributions from electric (blue) and magnetic multipoles (red) to the total SCS for the disk (e) and ring (f) with the inner/outer diameter of 800/1000 nm. The ring thickness is 80 nm, and the refractive index of surrounding is assumed to be $n = 1.48$. Magenta line indicates the wavelength, at which the scattering is dominated by electric octopole (c) or magnetic hexadecapole (f) contribution.
	}
  \label{fig2}
\end{figure*}

\section*{Scattering of disk and ring nanoparticles}

In order to test the generality of our findings, we replaced a sphere with its flat analog -- a disk -- and transformed it into a ring by introducing a void in its center (Figure~\ref{fig2}). It appeared that for the disk/ring structure multipole spectra does not have simple resonance shapes as for the sphere/shell particle, and an evolution of these spectra with a change of the void size is rather complicated (see Supporting Information, Figures~\ref{S4}-\ref{S7}). However, the general trend remains the same: with the increase of the hole the contributions from low-order multipoles decrease faster, compared to the high-order multipoles. Additionally, due to the symmetry and small thickness of the particle, the contributions from electric multipoles of the odd order and magnetic multipoles of the even order are negligible, simplifying the quest to find a regime with nearly pure high-order multipole scattering. We found that at the wavelength around $\lambda \approx 800$ nm and ring thickness of 80 nm, the scattering is dominated by the electric octopole (EO) for the ring with the inner/outer diameter of 400/800 nm (Figure~\ref{fig2}c), and it is dominated by the magnetic hexadecapole (MH) for the ring with the inner/outer diameter of 800/1000 nm (Figure~\ref{fig2}f). Here, the total scattering is calculated by numerical simulations (see Methods), from which multipole contributions are calculated, using the electric field inside the particle (in a same way as in our previous works \cite{anapole, slit}).

\begin{figure*}[htb]
\centering\includegraphics{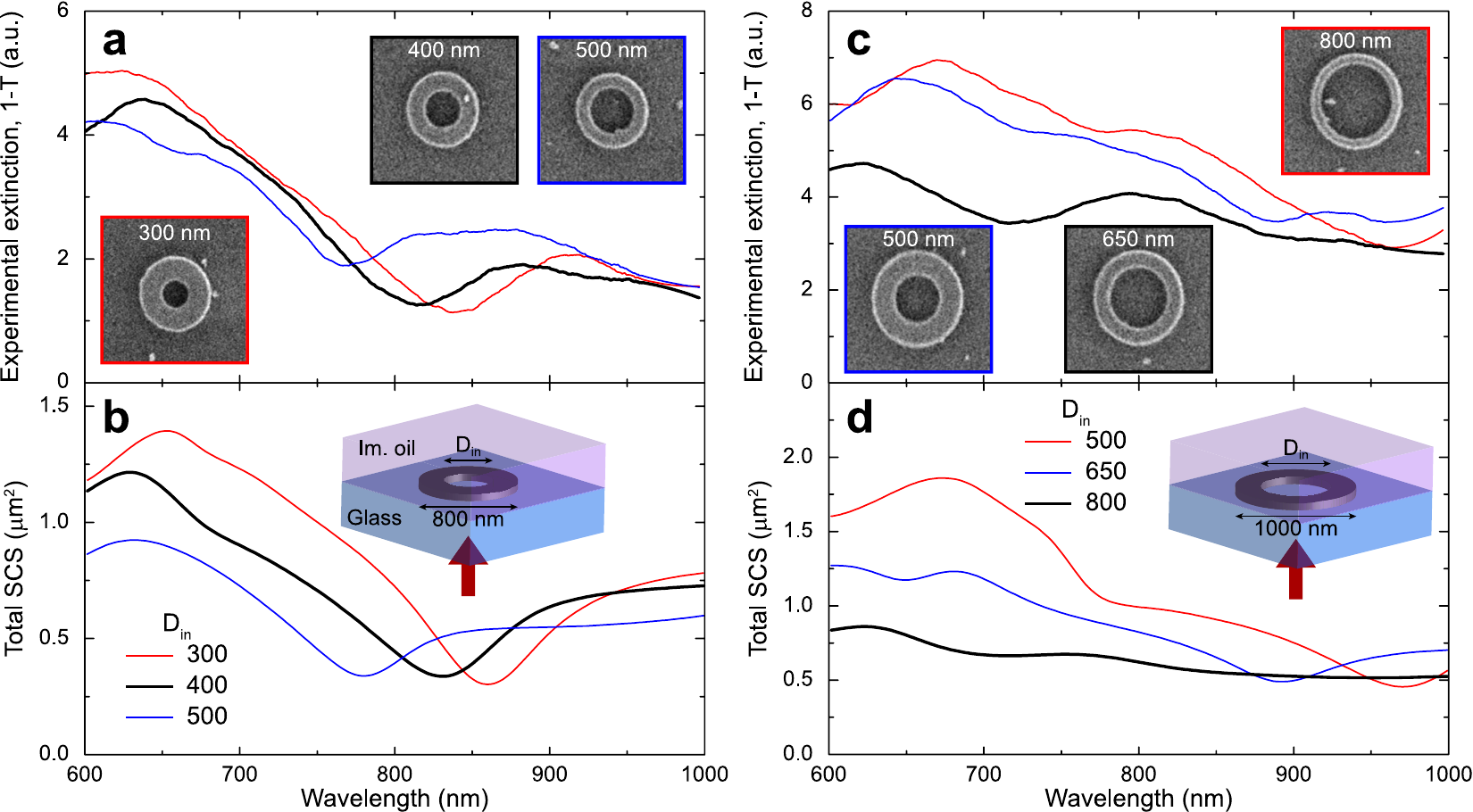}
  \caption{
(a,c) Experimental far-field extinction spectra ($1-T$) and (b,d) simulated scattering spectra of Si rings with varied internal hole diameter $D_{\rm in}$ and external diameter of 800 nm (a,b) and 1000 nm (c,d), respectively. In experiments the substrate is glass ($n \approx 1.45$), and the superstrate is immersion oil ($n = 1.518$). The rings are illuminated from the glass side (illustrated with red arrow). In simulations the ring is embedded in even surrounding with the refractive index of $n = 1.48$. Insets show SEM images of the structures, with labeled inner void diameter (panel size: 1500 nm).
}
  \label{fig3}
\end{figure*}

To confirm such selective scattering, we fabricated a series of isolated 80-nm-thick Si rings with varied inner (void) diameter and fixed outer diameter of 800 and 1000 nm. The fabrication was done by deposition of amorphous Si on a glass substrate, followed by etching through the mask (see Methods). First we measured the far-field transmission $T$ for each ring (see Methods), plot it as extinction $1-T$, and compared with simulated total scattering (Figure~\ref{fig3}).

One can already note from Figure~\ref{fig2} that the region with the dominant scattering by a single high-order multipole is not reflected in the total scattering spectrum. However, there are still some distinct features in spectra (dips and peaks), therefore by comparing them in measurements and simulations (Figure~\ref{fig3}), one can indirectly verify the correspondence between simulations and experiments.

\begin{figure*}[!htb]
\centering\includegraphics{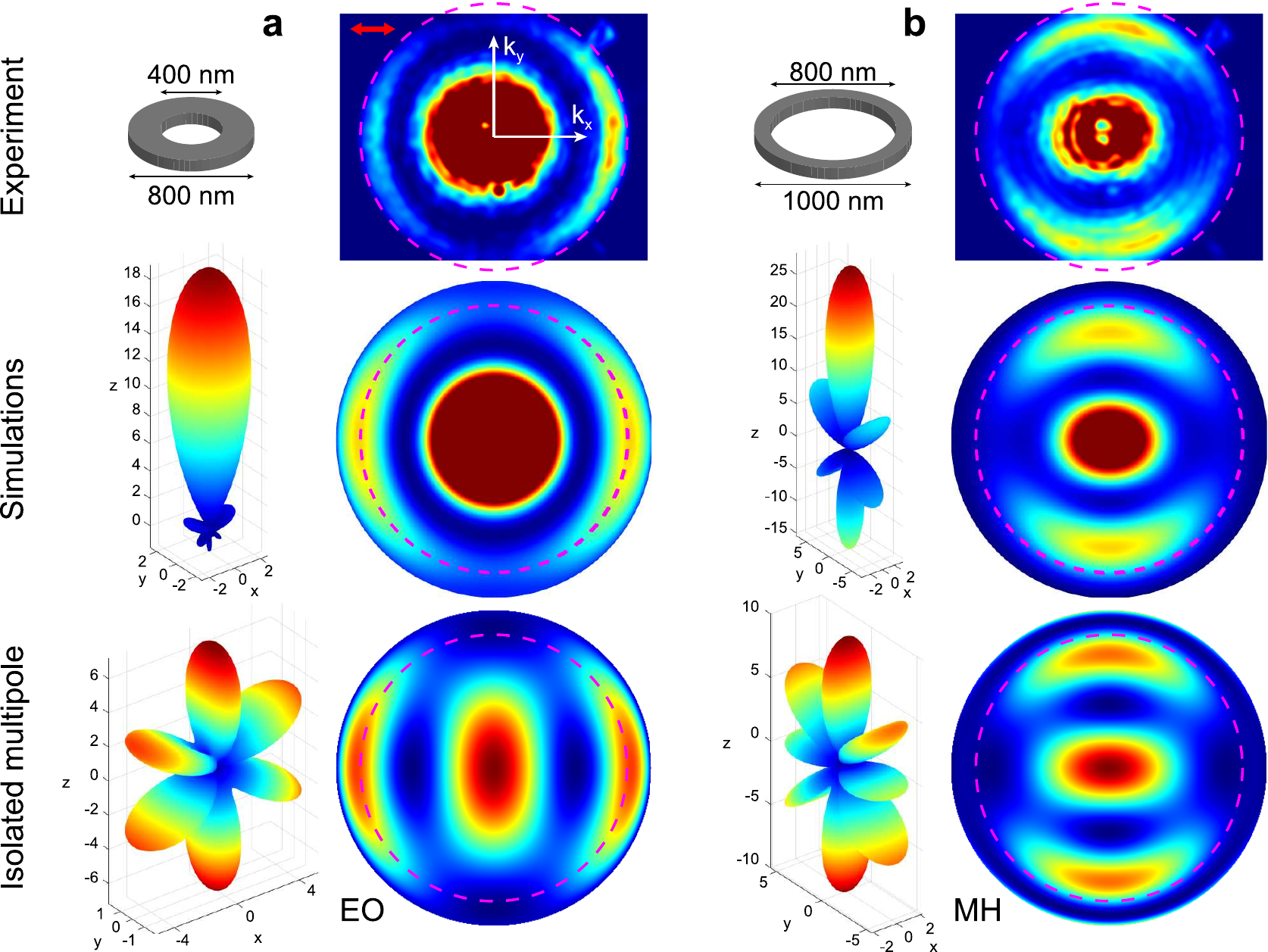}
  \caption{
Scattering diagrams of Si rings. (a,b) Experimentally measured and simulated scattering diagrams of Si rings with (a) 400/800 nm and (b) 800/1000 nm inner/outer diameter, respectively, compared with radiation diagrams of isolated (a) electric octopole (EO) and (b) magnetic hexadecapole (MH). The incident polarization is illustrated with a red arrow in (a). Magenta line in Fourier plane images shows the experimental limit for the collection angle and corresponds to the NA of the immersion-oil objective (1.25).
}
  \label{fig4}
\end{figure*}

Then we proceeded to direct measurements of the scattering diagram of Si rings (see Methods). The experimentally measured scattering diagrams of rings with 400/800 nm and 800/1000 nm inner/outer diameter are shown in Figure~\ref{fig4}, compared with simulated total scattering diagram and analytically calculated radiation pattern of an isolated multipole. Regarding the latter, one can directly calculate contributions of the selected multipoles to the scattering diagram \cite{Evlyukhin2019}, when their multipole moment is known (traceless and symmetrical tensors of rank 3 for octopole and rank 4 for hexadecapole). However, to simplify the analysis, we used direct expressions for the scattering diagrams of these multipoles for a perfect sphere in an even dielectric environment \cite{Barnes2010, Bohren}:
\begin{multline*}
{\sigma _{{\rm{EO}}}}\left( {\theta ,\varphi } \right) \propto {\cos ^2}\varphi {\left[ {\left( {5{{\cos }^2}\theta  - 1} \right)} \right]^2} \\
+ {\sin ^2}\varphi {\left[ {\cos \theta \left( {15{{\cos }^2}\theta  - 11} \right)} \right]^2},
\end{multline*}
\begin{multline*}
{\sigma _{{\rm{MH}}}}\left( {\theta ,\varphi } \right) \propto {\sin ^2}\varphi {\left[ {\cos \theta \left( {7{{\cos }^2}\theta  - 3} \right)} \right]^2} \\
+ {\cos ^2}\varphi {\left[ {\left( {28{{\cos }^4}\theta  - 27{{\cos }^2}\theta  + 3} \right)} \right]^2},
\end{multline*}
where $\theta$ and $\varphi$ are the polar and azimuthal angles, respectively, and $E$ ($M$) stands for electric (magnetic) multipole (see Supporting information, Supplementary Note 1). Though dependence of multipole contributions on the wavelength and void size for disk/ring structure is quite different from the one of the sphere/shell particle, we expect similar scattering diagrams for isolated multipoles of these particles due to the same mirror symmetry of the sphere and the ring. The above assumption and the validity of simulations in general are verified by a good agreement between experiment, simulations, and analytical calculations (Figure~\ref{fig4}). 

Moreover, the scattering diagrams have a well-pronounced feature -- scattering side-lobes along (Figure~\ref{fig4}a) or across the orientation of the incident beam polarization (Figure~\ref{fig4}b), which is a clear indication of the electric or magnetic multipole scattering. In case of the ring with the inner/outer diameter of 400/800 nm (Figure~\ref{fig4}a), scattering side-lobes are at around $\rm NA \approx 1.25$ in the Fourier plane, corresponding to the angle of $\rm arcsin(1.25/1.48) \approx \pi/3$, clearly confirming its electric octopole origin (see Supplementary Note 1). As for the other ring with the inner/outer diameter of 800/1000 nm (Figure~\ref{fig4}b), scattering side-lobes reach the maximum at around $\rm NA \approx 1$ in the Fourier plane, corresponding to the angle of $\rm arcsin(1/1.48) \approx \pi/4$, which is in accordance with the magnetic hexadecapole. Interestingly, the number of principal scattering lobes (6 for octopole and 8 for hexadecapole) cannot be anticipated from the multipole name, but it rather follows $2n$, where $n$ is the multipole order. This is not a contradiction, because the multipole name reflects the minimum number of oscillating point charges required to produce the given multipole moment, while it doesn`t state the number of scattering lobes. The deviation between scattering diagrams of each ring and the radiation pattern of the corresponding isolated multipole is due to the interference with other multipoles` contributions (EQ, MQ, MO, and MH for the ring with the inner/outer diameter of 400/800 nm, and ED, MQ, and EO for the ring with the inner/outer diameter of 800/1000 nm, as follows from Figure~\ref{fig2} at $\lambda = 800$ nm). This interference can lead to a strong suppression of some scattering lobes, but their angular positions are only slightly affected, which is an indication of the dominating single high-order multipole contribution (see Supplementary Information, Figure S9, for more discussion). The resulted asymmetry in forward/backward scattering is the so-called Mie effect \cite{Mie,Wolf}, which is a transition between symmetric Rayleigh scattering and diffraction.

\section*{Conclusion}

In summary, we demonstrated that the relative contribution of high-order multipoles can be boosted by introducing the void inside a high-refractive index nanoparticle. Using such a method allows finding a composition of the nanoparticle, whose scattering will be dominated by a single high-order multipole. We proved this method on a disk/ring shape, using Si as a high-refractive index material with a thickness of 80 nm. First we numerically found two structures producing nearly pure electric octopole and magnetic hexadecapole scattering. Then we fabricated such structures, directly measured their scattering diagrams, and finally verified the dominant high-order multipole contribution. Additionally to providing an intuitive understanding of the relation between shape and multipoles, our results can directly benefit applications in metasurfaces, sensing, quantum communications, and topological photonics.

\section*{Methods}

{\bf Fabrication.} The rings were fabricated by etching amorphous silicon, deposited on a silica wafer. First the fresh silica wafer was cleaned using a standard RCA clean, without the HF steps. Then 80 nm of Si was deposited using LPCVD. Standard reflectometry (FilmTek 4000TM) was used to measure Si thickness and refractive index (Supporting Information, Figure~\ref{S1}). After the Si deposition, AR-P 6200 resist from Allresist was spun at 200 nm, followed by thermal evaporation of 20 nm Al to be used as discharge path during the exposure. The structures were defined by electron-beam exposure. The next step involves Al layer removal in diluted phosphoric acid and development of the resist. The patterned resist is then used as a mask for etching the Si layer using a Bosch process and then removed using low power oxygen plasma.

{\bf Numerical simulations.} Scattering spectra, diagrams, and the electric field inside the structures (for multipole decomposition) were calculated using a 3D simulation with the finite-element method (FEM) implemented in a commercial software (COMSOL Multiphysics). A simulation sphere with a diameter of 2.4 $\mu$m was used with the perfectly-matched layer at the outer boundary and a tetrahedral mesh with a mesh size of $\sim$30 nm inside the silicon and $\sim$80 nm for the rest. The silicon ring was excited by a normal-incident CW plane wave, whose wavelength was swept to calculate spectra. For all calculations the permittivity value of Si was taken from measurements (Supporting Information, Figure~\ref{S1}), the refractive index of surrounding environment was set to 1.48.

{\bf Measurements of the scattering diagram.} Schematic diagram of the experimental setup for measuring scattering diagrams is shown in Supporting Information, Figure~\ref{S8}. The sample was illuminated using a linearly polarized Ti:Sapphire laser, tuned at a wavelength of $\sim$800 nm. The laser beam was weakly focused onto the sample using a 10$\times$ objective of numerical aperture NA = 0.20. The full-width-at-half-maximum (FHWM) of the focused beam spot was $\sim$5 $\mu$m. The scattered light was collected using a 63$\times$ oil-immersion collection objective, with a NA = 1.25. The structures on the sample were positioned facing the oil-immersion objective, embedded in the index-matching oil ($n = 1.518$). An imaging system, which consists of two lenses and two spatial filters, was used to image the back focal plane (BFP) of the collection objective with a charge-coupled device (CCD) camera. The BFP and its image are also referred to here as the Fourier plane, because it shows angular distribution of the scattering (i.e., scattering diagram). The two spatial filters have the following important functionalities. The first filter is a micrometric metallic ball (diameter $\sim$300 $\mu$m), glued on a glass substrate, and is used to stop the directly-transmitted light from reaching the CCD camera to avoid saturation. It is a Fourier-plane filter, and ideally it should be placed at the BFP inside the collection objective. Nevertheless, since the objective collimates the scattered light, the filter can be placed at the rear aperture of the objective and still produce the same filtering effect. The second filter is a pinhole, positioned at the image plane (focal point of the first lens), and is used to stop all the unwanted scattering from the surroundings of the nanostructure (i.e., from impurities in the glass substrate and oil). The images are captured with the CCD camera, located at the focal distance of the second lens, which matches with the Fourier plane.

{\bf Far-field spectroscopy.} Transmission spectroscopy was performed on a standard inverted optical microscope (Zeiss Axio Observer), equipped with a halogen light source, modified detection path, and fiber-coupled spectrometer (Ocean Optics QE Pro). The light was collected using a $\times$100 immersion-oil objective (Zeiss $\alpha$ Plan-FLUAR, NA = 1.45) and the same index-matching oil ($n = 1.518$). Schematically the setup is similar to the one for measuring scattering diagrams (Supporting Information, Figure~\ref{S8}), allowing filtering both in Fourier and direct image planes. We used an iris diaphragm for Fourier plane to limit detected NA to $\sim$0.3 (in order to replicate measurements of transmitted plane wave), and another iris diaphragm was set as a filter in the image plane, corresponding to the area with a diameter of $\sim$3 $\mu$m in a sample plane (that is, when image plane was recorded by a camera, the opening was $\sim$3 times larger than the ring with the outer diameter of 1 $\mu$m). The measured transmission spectra were normalized on the transmission spectrum through the glass substrate without any structure.

\textbf{Acknowledgments}

The authors acknowledge financial support from the European Research Council (the PLAQNAP project, Grant No. 341054) and the University of Southern Denmark (SDU2020 funding), from scholarship 299967. N.A.M. is a VILLUM Investigator supported by Villum Fonden (Grant No. 16498). C.E.G.-O and V.C. acknowledge the technical assistance of Fabiola Armenta with the experimental setup. V.C. and C.E.G.-O. acknowledge funding from CONACYT Basic Scientific Research Grants No. 250719 and No. 252621. RM and AVL acknowledge the financial support from Villum Fonden ''DarkSILD project'' (Grant No. 11116) as well as the support of the National Centre for Nano Fabrication and Characterization (DTU Nanolab) for fabrication of the structures. A.B.E. and B.N.C. acknowledge financial support from the Deutsche Forschungsgemeinschaft (DFG, German Research Foundation) under Germany's Excellence Strategy within the Cluster of Excellence PhoenixD (EXC 2122, Project No. 390833453) and DFG Project CH179/34-1. Numerical simulation was partially supported by the Russian Science Foundation (Grant No. 18-19-00684).

\clearpage

\bibliography{zenin}

\begin{thebibliography}{44}%
\makeatletter
\providecommand \@ifxundefined [1]{%
 \@ifx{#1\undefined}
}%
\providecommand \@ifnum [1]{%
 \ifnum #1\expandafter \@firstoftwo
 \else \expandafter \@secondoftwo
 \fi
}%
\providecommand \@ifx [1]{%
 \ifx #1\expandafter \@firstoftwo
 \else \expandafter \@secondoftwo
 \fi
}%
\providecommand \natexlab [1]{#1}%
\providecommand \enquote  [1]{``#1''}%
\providecommand \bibnamefont  [1]{#1}%
\providecommand \bibfnamefont [1]{#1}%
\providecommand \citenamefont [1]{#1}%
\providecommand \href@noop [0]{\@secondoftwo}%
\providecommand \href [0]{\begingroup \@sanitize@url \@href}%
\providecommand \@href[1]{\@@startlink{#1}\@@href}%
\providecommand \@@href[1]{\endgroup#1\@@endlink}%
\providecommand \@sanitize@url [0]{\catcode `\\12\catcode `\$12\catcode
  `\&12\catcode `\#12\catcode `\^12\catcode `\_12\catcode `\%12\relax}%
\providecommand \@@startlink[1]{}%
\providecommand \@@endlink[0]{}%
\providecommand \url  [0]{\begingroup\@sanitize@url \@url }%
\providecommand \@url [1]{\endgroup\@href {#1}{\urlprefix }}%
\providecommand \urlprefix  [0]{URL }%
\providecommand \Eprint [0]{\href }%
\providecommand \doibase [0]{http://dx.doi.org/}%
\providecommand \selectlanguage [0]{\@gobble}%
\providecommand \bibinfo  [0]{\@secondoftwo}%
\providecommand \bibfield  [0]{\@secondoftwo}%
\providecommand \translation [1]{[#1]}%
\providecommand \BibitemOpen [0]{}%
\providecommand \bibitemStop [0]{}%
\providecommand \bibitemNoStop [0]{.\EOS\space}%
\providecommand \EOS [0]{\spacefactor3000\relax}%
\providecommand \BibitemShut  [1]{\csname bibitem#1\endcsname}%
\let\auto@bib@innerbib\@empty
\bibitem [{\citenamefont {Evlyukhin}\ \emph {et~al.}(2010)\citenamefont
  {Evlyukhin}, \citenamefont {Reinhardt}, \citenamefont {Seidel}, \citenamefont
  {Luk{\textquoteright}yanchuk},\ and\ \citenamefont
  {Chichkov}}]{Evlyukhin2010}%
  \BibitemOpen
  \bibfield  {author} {\bibinfo {author} {\bibfnamefont {A.~B.}\ \bibnamefont
  {Evlyukhin}}, \bibinfo {author} {\bibfnamefont {C.}~\bibnamefont
  {Reinhardt}}, \bibinfo {author} {\bibfnamefont {A.}~\bibnamefont {Seidel}},
  \bibinfo {author} {\bibfnamefont {B.~S.}\ \bibnamefont
  {Luk{\textquoteright}yanchuk}}, \ and\ \bibinfo {author} {\bibfnamefont
  {B.~N.}\ \bibnamefont {Chichkov}},\ }\href {\doibase
  10.1103/PhysRevB.82.045404} {\bibfield  {journal} {\bibinfo  {journal} {Phys.
  Rev. B}\ }\textbf {\bibinfo {volume} {82}},\ \bibinfo {pages} {045404}
  (\bibinfo {year} {2010})}\BibitemShut {NoStop}%
\bibitem [{\citenamefont {Evlyukhin}\ \emph {et~al.}(2012)\citenamefont
  {Evlyukhin}, \citenamefont {Novikov}, \citenamefont {Zywietz}, \citenamefont
  {Eriksen}, \citenamefont {Reinhardt}, \citenamefont {Bozhevolnyi},\ and\
  \citenamefont {Chichkov}}]{NL2012}%
  \BibitemOpen
  \bibfield  {author} {\bibinfo {author} {\bibfnamefont {A.~B.}\ \bibnamefont
  {Evlyukhin}}, \bibinfo {author} {\bibfnamefont {S.~M.}\ \bibnamefont
  {Novikov}}, \bibinfo {author} {\bibfnamefont {U.}~\bibnamefont {Zywietz}},
  \bibinfo {author} {\bibfnamefont {R.~L.}\ \bibnamefont {Eriksen}}, \bibinfo
  {author} {\bibfnamefont {C.}~\bibnamefont {Reinhardt}}, \bibinfo {author}
  {\bibfnamefont {S.~I.}\ \bibnamefont {Bozhevolnyi}}, \ and\ \bibinfo {author}
  {\bibfnamefont {B.~N.}\ \bibnamefont {Chichkov}},\ }\href {\doibase
  10.1021/nl301594s} {\bibfield  {journal} {\bibinfo  {journal} {Nano Letters}\
  }\textbf {\bibinfo {volume} {12}},\ \bibinfo {pages} {3749} (\bibinfo {year}
  {2012})}\BibitemShut {NoStop}%
\bibitem [{\citenamefont {Kuznetsov}\ \emph {et~al.}(2016)\citenamefont
  {Kuznetsov}, \citenamefont {Miroshnichenko}, \citenamefont {Brongersma},
  \citenamefont {Kivshar},\ and\ \citenamefont
  {Luk{\textquoteright}yanchuk}}]{diel1}%
  \BibitemOpen
  \bibfield  {author} {\bibinfo {author} {\bibfnamefont {A.~I.}\ \bibnamefont
  {Kuznetsov}}, \bibinfo {author} {\bibfnamefont {A.~E.}\ \bibnamefont
  {Miroshnichenko}}, \bibinfo {author} {\bibfnamefont {M.~L.}\ \bibnamefont
  {Brongersma}}, \bibinfo {author} {\bibfnamefont {Y.~S.}\ \bibnamefont
  {Kivshar}}, \ and\ \bibinfo {author} {\bibfnamefont {B.}~\bibnamefont
  {Luk{\textquoteright}yanchuk}},\ }\href {\doibase 10.1126/science.aag2472}
  {\bibfield  {journal} {\bibinfo  {journal} {Science}\ }\textbf {\bibinfo
  {volume} {354}},\ \bibinfo {pages} {aag2472} (\bibinfo {year}
  {2016})}\BibitemShut {NoStop}%
\bibitem [{\citenamefont {Staude}\ and\ \citenamefont
  {Schilling}(2017)}]{diel2}%
  \BibitemOpen
  \bibfield  {author} {\bibinfo {author} {\bibfnamefont {I.}~\bibnamefont
  {Staude}}\ and\ \bibinfo {author} {\bibfnamefont {J.}~\bibnamefont
  {Schilling}},\ }\href {\doibase 10.1038/nphoton.2017.39} {\bibfield
  {journal} {\bibinfo  {journal} {Nat. Photon.}\ }\textbf {\bibinfo {volume}
  {11}},\ \bibinfo {pages} {274} (\bibinfo {year} {2017})}\BibitemShut
  {NoStop}%
\bibitem [{\citenamefont {Baranov}\ \emph {et~al.}(2017)\citenamefont
  {Baranov}, \citenamefont {Zuev}, \citenamefont {Lepeshov}, \citenamefont
  {Kotov}, \citenamefont {Krasnok}, \citenamefont {Evlyukhin},\ and\
  \citenamefont {Chichkov}}]{fabr}%
  \BibitemOpen
  \bibfield  {author} {\bibinfo {author} {\bibfnamefont {D.~G.}\ \bibnamefont
  {Baranov}}, \bibinfo {author} {\bibfnamefont {D.~A.}\ \bibnamefont {Zuev}},
  \bibinfo {author} {\bibfnamefont {S.~I.}\ \bibnamefont {Lepeshov}}, \bibinfo
  {author} {\bibfnamefont {O.~V.}\ \bibnamefont {Kotov}}, \bibinfo {author}
  {\bibfnamefont {A.~E.}\ \bibnamefont {Krasnok}}, \bibinfo {author}
  {\bibfnamefont {A.~B.}\ \bibnamefont {Evlyukhin}}, \ and\ \bibinfo {author}
  {\bibfnamefont {B.~N.}\ \bibnamefont {Chichkov}},\ }\href {\doibase
  10.1364/OPTICA.4.000814} {\bibfield  {journal} {\bibinfo  {journal} {Optica}\
  }\textbf {\bibinfo {volume} {4}},\ \bibinfo {pages} {814} (\bibinfo {year}
  {2017})}\BibitemShut {NoStop}%
\bibitem [{\citenamefont {Shibanuma}\ \emph {et~al.}(2017)\citenamefont
  {Shibanuma}, \citenamefont {Matsui}, \citenamefont {Roschuk}, \citenamefont
  {Wojcik}, \citenamefont {Mascher}, \citenamefont {Albella},\ and\
  \citenamefont {Maier}}]{meta1}%
  \BibitemOpen
  \bibfield  {author} {\bibinfo {author} {\bibfnamefont {T.}~\bibnamefont
  {Shibanuma}}, \bibinfo {author} {\bibfnamefont {T.}~\bibnamefont {Matsui}},
  \bibinfo {author} {\bibfnamefont {T.}~\bibnamefont {Roschuk}}, \bibinfo
  {author} {\bibfnamefont {J.}~\bibnamefont {Wojcik}}, \bibinfo {author}
  {\bibfnamefont {P.}~\bibnamefont {Mascher}}, \bibinfo {author} {\bibfnamefont
  {P.}~\bibnamefont {Albella}}, \ and\ \bibinfo {author} {\bibfnamefont
  {S.~A.}\ \bibnamefont {Maier}},\ }\href {\doibase
  10.1021/acsphotonics.6b00979} {\bibfield  {journal} {\bibinfo  {journal} {ACS
  Photonics}\ }\textbf {\bibinfo {volume} {4}},\ \bibinfo {pages} {489}
  (\bibinfo {year} {2017})}\BibitemShut {NoStop}%
\bibitem [{\citenamefont {Khaidarov}\ \emph {et~al.}(2017)\citenamefont
  {Khaidarov}, \citenamefont {Hao}, \citenamefont {Paniagua-Dom{\'i}nguez},
  \citenamefont {Yu}, \citenamefont {Fu}, \citenamefont {Valuckas},
  \citenamefont {Yap}, \citenamefont {Toh}, \citenamefont {Ng},\ and\
  \citenamefont {Kuznetsov}}]{meta2}%
  \BibitemOpen
  \bibfield  {author} {\bibinfo {author} {\bibfnamefont {E.}~\bibnamefont
  {Khaidarov}}, \bibinfo {author} {\bibfnamefont {H.}~\bibnamefont {Hao}},
  \bibinfo {author} {\bibfnamefont {R.}~\bibnamefont {Paniagua-Dom{\'i}nguez}},
  \bibinfo {author} {\bibfnamefont {Y.~F.}\ \bibnamefont {Yu}}, \bibinfo
  {author} {\bibfnamefont {Y.~H.}\ \bibnamefont {Fu}}, \bibinfo {author}
  {\bibfnamefont {V.}~\bibnamefont {Valuckas}}, \bibinfo {author}
  {\bibfnamefont {S.~L.~K.}\ \bibnamefont {Yap}}, \bibinfo {author}
  {\bibfnamefont {Y.~T.}\ \bibnamefont {Toh}}, \bibinfo {author} {\bibfnamefont
  {J.~S.~K.}\ \bibnamefont {Ng}}, \ and\ \bibinfo {author} {\bibfnamefont
  {A.~I.}\ \bibnamefont {Kuznetsov}},\ }\href {\doibase
  10.1021/acs.nanolett.7b02952} {\bibfield  {journal} {\bibinfo  {journal}
  {Nano Letters}\ }\textbf {\bibinfo {volume} {17}},\ \bibinfo {pages} {6267}
  (\bibinfo {year} {2017})}\BibitemShut {NoStop}%
\bibitem [{\citenamefont {Paniagua-Dom{\'i}nguez}\ \emph
  {et~al.}(2018)\citenamefont {Paniagua-Dom{\'i}nguez}, \citenamefont {Yu},
  \citenamefont {Khaidarov}, \citenamefont {Choi}, \citenamefont {Leong},
  \citenamefont {Bakker}, \citenamefont {Liang}, \citenamefont {Fu},
  \citenamefont {Valuckas}, \citenamefont {Krivitsky},\ and\ \citenamefont
  {Kuznetsov}}]{meta3}%
  \BibitemOpen
  \bibfield  {author} {\bibinfo {author} {\bibfnamefont {R.}~\bibnamefont
  {Paniagua-Dom{\'i}nguez}}, \bibinfo {author} {\bibfnamefont {Y.~F.}\
  \bibnamefont {Yu}}, \bibinfo {author} {\bibfnamefont {E.}~\bibnamefont
  {Khaidarov}}, \bibinfo {author} {\bibfnamefont {S.}~\bibnamefont {Choi}},
  \bibinfo {author} {\bibfnamefont {V.}~\bibnamefont {Leong}}, \bibinfo
  {author} {\bibfnamefont {R.~M.}\ \bibnamefont {Bakker}}, \bibinfo {author}
  {\bibfnamefont {X.}~\bibnamefont {Liang}}, \bibinfo {author} {\bibfnamefont
  {Y.~H.}\ \bibnamefont {Fu}}, \bibinfo {author} {\bibfnamefont
  {V.}~\bibnamefont {Valuckas}}, \bibinfo {author} {\bibfnamefont {L.~A.}\
  \bibnamefont {Krivitsky}}, \ and\ \bibinfo {author} {\bibfnamefont {A.~I.}\
  \bibnamefont {Kuznetsov}},\ }\href {\doibase 10.1021/acs.nanolett.8b00368}
  {\bibfield  {journal} {\bibinfo  {journal} {Nano Letters}\ }\textbf {\bibinfo
  {volume} {18}},\ \bibinfo {pages} {2124} (\bibinfo {year}
  {2018})}\BibitemShut {NoStop}%
\bibitem [{\citenamefont {Bucher}\ \emph {et~al.}(2019)\citenamefont {Bucher},
  \citenamefont {Vaskin}, \citenamefont {Mupparapu}, \citenamefont
  {L{\"o}chner}, \citenamefont {George}, \citenamefont {Chong}, \citenamefont
  {Fasold}, \citenamefont {Neumann}, \citenamefont {Choi}, \citenamefont
  {Eilenberger}, \citenamefont {Setzpfandt}, \citenamefont {Kivshar},
  \citenamefont {Pertsch}, \citenamefont {Turchanin},\ and\ \citenamefont
  {Staude}}]{meta4}%
  \BibitemOpen
  \bibfield  {author} {\bibinfo {author} {\bibfnamefont {T.}~\bibnamefont
  {Bucher}}, \bibinfo {author} {\bibfnamefont {A.}~\bibnamefont {Vaskin}},
  \bibinfo {author} {\bibfnamefont {R.}~\bibnamefont {Mupparapu}}, \bibinfo
  {author} {\bibfnamefont {F.~J.~F.}\ \bibnamefont {L{\"o}chner}}, \bibinfo
  {author} {\bibfnamefont {A.}~\bibnamefont {George}}, \bibinfo {author}
  {\bibfnamefont {K.~E.}\ \bibnamefont {Chong}}, \bibinfo {author}
  {\bibfnamefont {S.}~\bibnamefont {Fasold}}, \bibinfo {author} {\bibfnamefont
  {C.}~\bibnamefont {Neumann}}, \bibinfo {author} {\bibfnamefont {D.-Y.}\
  \bibnamefont {Choi}}, \bibinfo {author} {\bibfnamefont {F.}~\bibnamefont
  {Eilenberger}}, \bibinfo {author} {\bibfnamefont {F.}~\bibnamefont
  {Setzpfandt}}, \bibinfo {author} {\bibfnamefont {Y.~S.}\ \bibnamefont
  {Kivshar}}, \bibinfo {author} {\bibfnamefont {T.}~\bibnamefont {Pertsch}},
  \bibinfo {author} {\bibfnamefont {A.}~\bibnamefont {Turchanin}}, \ and\
  \bibinfo {author} {\bibfnamefont {I.}~\bibnamefont {Staude}},\ }\href
  {\doibase 10.1021/acsphotonics.8b01771} {\bibfield  {journal} {\bibinfo
  {journal} {ACS Photonics}\ }\textbf {\bibinfo {volume} {6}},\ \bibinfo
  {pages} {1002} (\bibinfo {year} {2019})}\BibitemShut {NoStop}%
\bibitem [{\citenamefont {Kruk}\ \emph {et~al.}(2018)\citenamefont {Kruk},
  \citenamefont {Ferreira}, \citenamefont {Mac~Suibhne}, \citenamefont
  {Tsekrekos}, \citenamefont {Kravchenko}, \citenamefont {Ellis}, \citenamefont
  {Neshev}, \citenamefont {Turitsyn},\ and\ \citenamefont {Kivshar}}]{meta5}%
  \BibitemOpen
  \bibfield  {author} {\bibinfo {author} {\bibfnamefont {S.}~\bibnamefont
  {Kruk}}, \bibinfo {author} {\bibfnamefont {F.}~\bibnamefont {Ferreira}},
  \bibinfo {author} {\bibfnamefont {N.}~\bibnamefont {Mac~Suibhne}}, \bibinfo
  {author} {\bibfnamefont {C.}~\bibnamefont {Tsekrekos}}, \bibinfo {author}
  {\bibfnamefont {I.}~\bibnamefont {Kravchenko}}, \bibinfo {author}
  {\bibfnamefont {A.}~\bibnamefont {Ellis}}, \bibinfo {author} {\bibfnamefont
  {D.}~\bibnamefont {Neshev}}, \bibinfo {author} {\bibfnamefont
  {S.}~\bibnamefont {Turitsyn}}, \ and\ \bibinfo {author} {\bibfnamefont
  {Y.}~\bibnamefont {Kivshar}},\ }\href {\doibase 10.1002/lpor.201800031}
  {\bibfield  {journal} {\bibinfo  {journal} {Laser \& Photonics Reviews}\
  }\textbf {\bibinfo {volume} {12}},\ \bibinfo {pages} {1800031} (\bibinfo
  {year} {2018})}\BibitemShut {NoStop}%
\bibitem [{\citenamefont {Tang}\ \emph {et~al.}(2019)\citenamefont {Tang},
  \citenamefont {Li}, \citenamefont {Pan}, \citenamefont {Zhou}, \citenamefont
  {Jiang},\ and\ \citenamefont {Ding}}]{meta6}%
  \BibitemOpen
  \bibfield  {author} {\bibinfo {author} {\bibfnamefont {S.}~\bibnamefont
  {Tang}}, \bibinfo {author} {\bibfnamefont {X.}~\bibnamefont {Li}}, \bibinfo
  {author} {\bibfnamefont {W.}~\bibnamefont {Pan}}, \bibinfo {author}
  {\bibfnamefont {J.}~\bibnamefont {Zhou}}, \bibinfo {author} {\bibfnamefont
  {T.}~\bibnamefont {Jiang}}, \ and\ \bibinfo {author} {\bibfnamefont
  {F.}~\bibnamefont {Ding}},\ }\href {\doibase 10.1364/OE.27.004281} {\bibfield
   {journal} {\bibinfo  {journal} {Opt. Express}\ }\textbf {\bibinfo {volume}
  {27}},\ \bibinfo {pages} {4281} (\bibinfo {year} {2019})}\BibitemShut
  {NoStop}%
\bibitem [{\citenamefont {Rahimzadegan}\ \emph {et~al.}(2019)\citenamefont
  {Rahimzadegan}, \citenamefont {Arslan}, \citenamefont {Dams}, \citenamefont
  {Groner}, \citenamefont {Garcia-Santiago}, \citenamefont {Alaee},
  \citenamefont {Fernandez-Corbaton}, \citenamefont {Pertsch}, \citenamefont
  {Staude},\ and\ \citenamefont {Rockstuhl}}]{metaR}%
  \BibitemOpen
  \bibfield  {author} {\bibinfo {author} {\bibfnamefont {A.}~\bibnamefont
  {Rahimzadegan}}, \bibinfo {author} {\bibfnamefont {D.}~\bibnamefont
  {Arslan}}, \bibinfo {author} {\bibfnamefont {D.}~\bibnamefont {Dams}},
  \bibinfo {author} {\bibfnamefont {A.}~\bibnamefont {Groner}}, \bibinfo
  {author} {\bibfnamefont {X.}~\bibnamefont {Garcia-Santiago}}, \bibinfo
  {author} {\bibfnamefont {R.}~\bibnamefont {Alaee}}, \bibinfo {author}
  {\bibfnamefont {I.}~\bibnamefont {Fernandez-Corbaton}}, \bibinfo {author}
  {\bibfnamefont {T.}~\bibnamefont {Pertsch}}, \bibinfo {author} {\bibfnamefont
  {I.}~\bibnamefont {Staude}}, \ and\ \bibinfo {author} {\bibfnamefont
  {C.}~\bibnamefont {Rockstuhl}},\ }\href {\doibase 10.1515/nanoph-2019-0239}
  {\bibfield  {journal} {\bibinfo  {journal} {Nanophotonics}\ ,\ \bibinfo
  {pages} {aop}} (\bibinfo {year} {2019})}\BibitemShut {NoStop}%
\bibitem [{\citenamefont {Dong}\ \emph {et~al.}(2017)\citenamefont {Dong},
  \citenamefont {Ho}, \citenamefont {Yu}, \citenamefont {Fu}, \citenamefont
  {Paniagua-Dom{\'i}nguez}, \citenamefont {Wang}, \citenamefont {Kuznetsov},\
  and\ \citenamefont {Yang}}]{color1}%
  \BibitemOpen
  \bibfield  {author} {\bibinfo {author} {\bibfnamefont {Z.}~\bibnamefont
  {Dong}}, \bibinfo {author} {\bibfnamefont {J.}~\bibnamefont {Ho}}, \bibinfo
  {author} {\bibfnamefont {Y.~F.}\ \bibnamefont {Yu}}, \bibinfo {author}
  {\bibfnamefont {Y.~H.}\ \bibnamefont {Fu}}, \bibinfo {author} {\bibfnamefont
  {R.}~\bibnamefont {Paniagua-Dom{\'i}nguez}}, \bibinfo {author} {\bibfnamefont
  {S.}~\bibnamefont {Wang}}, \bibinfo {author} {\bibfnamefont {A.~I.}\
  \bibnamefont {Kuznetsov}}, \ and\ \bibinfo {author} {\bibfnamefont
  {J.~K.~W.}\ \bibnamefont {Yang}},\ }\href {\doibase
  10.1021/acs.nanolett.7b03613} {\bibfield  {journal} {\bibinfo  {journal}
  {Nano Letters}\ }\textbf {\bibinfo {volume} {17}},\ \bibinfo {pages} {7620}
  (\bibinfo {year} {2017})}\BibitemShut {NoStop}%
\bibitem [{\citenamefont {Zhu}\ \emph {et~al.}(2017)\citenamefont {Zhu},
  \citenamefont {Yan}, \citenamefont {Levy}, \citenamefont {Mortensen},\ and\
  \citenamefont {Kristensen}}]{color2}%
  \BibitemOpen
  \bibfield  {author} {\bibinfo {author} {\bibfnamefont {X.}~\bibnamefont
  {Zhu}}, \bibinfo {author} {\bibfnamefont {W.}~\bibnamefont {Yan}}, \bibinfo
  {author} {\bibfnamefont {U.}~\bibnamefont {Levy}}, \bibinfo {author}
  {\bibfnamefont {N.~A.}\ \bibnamefont {Mortensen}}, \ and\ \bibinfo {author}
  {\bibfnamefont {A.}~\bibnamefont {Kristensen}},\ }\href {\doibase
  10.1126/sciadv.1602487} {\bibfield  {journal} {\bibinfo  {journal} {Science
  Advances}\ }\textbf {\bibinfo {volume} {3}} (\bibinfo {year} {2017}),\
  10.1126/sciadv.1602487}\BibitemShut {NoStop}%
\bibitem [{\citenamefont {Ha}\ \emph {et~al.}(2018)\citenamefont {Ha},
  \citenamefont {Fu}, \citenamefont {Emani}, \citenamefont {Pan}, \citenamefont
  {Bakker}, \citenamefont {Paniagua-Dom{\'i}nguez},\ and\ \citenamefont
  {Kuznetsov}}]{lasers1}%
  \BibitemOpen
  \bibfield  {author} {\bibinfo {author} {\bibfnamefont {S.~T.}\ \bibnamefont
  {Ha}}, \bibinfo {author} {\bibfnamefont {Y.~H.}\ \bibnamefont {Fu}}, \bibinfo
  {author} {\bibfnamefont {N.~K.}\ \bibnamefont {Emani}}, \bibinfo {author}
  {\bibfnamefont {Z.}~\bibnamefont {Pan}}, \bibinfo {author} {\bibfnamefont
  {R.~M.}\ \bibnamefont {Bakker}}, \bibinfo {author} {\bibfnamefont
  {R.}~\bibnamefont {Paniagua-Dom{\'i}nguez}}, \ and\ \bibinfo {author}
  {\bibfnamefont {A.~I.}\ \bibnamefont {Kuznetsov}},\ }\href {\doibase
  10.1038/s41565-018-0245-5} {\bibfield  {journal} {\bibinfo  {journal} {Nature
  Nanotechnology}\ }\textbf {\bibinfo {volume} {13}},\ \bibinfo {pages} {1042}
  (\bibinfo {year} {2018})}\BibitemShut {NoStop}%
\bibitem [{\citenamefont {Tiguntseva}\ \emph {et~al.}(2019)\citenamefont
  {Tiguntseva}, \citenamefont {Koshelev}, \citenamefont {Furasova},
  \citenamefont {Mikhailovskii}, \citenamefont {Ushakova}, \citenamefont
  {Baranov}, \citenamefont {Shegai}, \citenamefont {Zakhidov}, \citenamefont
  {Kivshar},\ and\ \citenamefont {Makarov}}]{lasers2}%
  \BibitemOpen
  \bibfield  {author} {\bibinfo {author} {\bibfnamefont {E.~Y.}\ \bibnamefont
  {Tiguntseva}}, \bibinfo {author} {\bibfnamefont {K.~L.}\ \bibnamefont
  {Koshelev}}, \bibinfo {author} {\bibfnamefont {A.~D.}\ \bibnamefont
  {Furasova}}, \bibinfo {author} {\bibfnamefont {V.~Y.}\ \bibnamefont
  {Mikhailovskii}}, \bibinfo {author} {\bibfnamefont {E.~V.}\ \bibnamefont
  {Ushakova}}, \bibinfo {author} {\bibfnamefont {D.~G.}\ \bibnamefont
  {Baranov}}, \bibinfo {author} {\bibfnamefont {T.~O.}\ \bibnamefont {Shegai}},
  \bibinfo {author} {\bibfnamefont {A.~A.}\ \bibnamefont {Zakhidov}}, \bibinfo
  {author} {\bibfnamefont {Y.~S.}\ \bibnamefont {Kivshar}}, \ and\ \bibinfo
  {author} {\bibfnamefont {S.~V.}\ \bibnamefont {Makarov}},\ }\href@noop {}
  {\enquote {\bibinfo {title} {Single-particle mie-resonant all-dielectric
  nanolasers},}\ } (\bibinfo {year} {2019})\BibitemShut {NoStop}%
\bibitem [{\citenamefont {Yesilkoy}\ \emph {et~al.}(2019)\citenamefont
  {Yesilkoy}, \citenamefont {Arvelo}, \citenamefont {Jahani}, \citenamefont
  {Liu}, \citenamefont {Tittl}, \citenamefont {Cevher}, \citenamefont
  {Kivshar},\ and\ \citenamefont {Altug}}]{bio1}%
  \BibitemOpen
  \bibfield  {author} {\bibinfo {author} {\bibfnamefont {F.}~\bibnamefont
  {Yesilkoy}}, \bibinfo {author} {\bibfnamefont {E.~R.}\ \bibnamefont
  {Arvelo}}, \bibinfo {author} {\bibfnamefont {Y.}~\bibnamefont {Jahani}},
  \bibinfo {author} {\bibfnamefont {M.}~\bibnamefont {Liu}}, \bibinfo {author}
  {\bibfnamefont {A.}~\bibnamefont {Tittl}}, \bibinfo {author} {\bibfnamefont
  {V.}~\bibnamefont {Cevher}}, \bibinfo {author} {\bibfnamefont
  {Y.}~\bibnamefont {Kivshar}}, \ and\ \bibinfo {author} {\bibfnamefont
  {H.}~\bibnamefont {Altug}},\ }\href {\doibase 10.1038/s41566-019-0394-6}
  {\bibfield  {journal} {\bibinfo  {journal} {Nature Photonics}\ }\textbf
  {\bibinfo {volume} {13}},\ \bibinfo {pages} {390} (\bibinfo {year}
  {2019})}\BibitemShut {NoStop}%
\bibitem [{\citenamefont {Leitis}\ \emph {et~al.}(2019)\citenamefont {Leitis},
  \citenamefont {Tittl}, \citenamefont {Liu}, \citenamefont {Lee},
  \citenamefont {Gu}, \citenamefont {Kivshar},\ and\ \citenamefont
  {Altug}}]{bio2}%
  \BibitemOpen
  \bibfield  {author} {\bibinfo {author} {\bibfnamefont {A.}~\bibnamefont
  {Leitis}}, \bibinfo {author} {\bibfnamefont {A.}~\bibnamefont {Tittl}},
  \bibinfo {author} {\bibfnamefont {M.}~\bibnamefont {Liu}}, \bibinfo {author}
  {\bibfnamefont {B.~H.}\ \bibnamefont {Lee}}, \bibinfo {author} {\bibfnamefont
  {M.~B.}\ \bibnamefont {Gu}}, \bibinfo {author} {\bibfnamefont {Y.~S.}\
  \bibnamefont {Kivshar}}, \ and\ \bibinfo {author} {\bibfnamefont
  {H.}~\bibnamefont {Altug}},\ }\href {\doibase 10.1126/sciadv.aaw2871}
  {\bibfield  {journal} {\bibinfo  {journal} {Science Advances}\ }\textbf
  {\bibinfo {volume} {5}} (\bibinfo {year} {2019}),\
  10.1126/sciadv.aaw2871}\BibitemShut {NoStop}%
\bibitem [{\citenamefont {Tittl}\ \emph {et~al.}(2018)\citenamefont {Tittl},
  \citenamefont {Leitis}, \citenamefont {Liu}, \citenamefont {Yesilkoy},
  \citenamefont {Choi}, \citenamefont {Neshev}, \citenamefont {Kivshar},\ and\
  \citenamefont {Altug}}]{bio3}%
  \BibitemOpen
  \bibfield  {author} {\bibinfo {author} {\bibfnamefont {A.}~\bibnamefont
  {Tittl}}, \bibinfo {author} {\bibfnamefont {A.}~\bibnamefont {Leitis}},
  \bibinfo {author} {\bibfnamefont {M.}~\bibnamefont {Liu}}, \bibinfo {author}
  {\bibfnamefont {F.}~\bibnamefont {Yesilkoy}}, \bibinfo {author}
  {\bibfnamefont {D.-Y.}\ \bibnamefont {Choi}}, \bibinfo {author}
  {\bibfnamefont {D.~N.}\ \bibnamefont {Neshev}}, \bibinfo {author}
  {\bibfnamefont {Y.~S.}\ \bibnamefont {Kivshar}}, \ and\ \bibinfo {author}
  {\bibfnamefont {H.}~\bibnamefont {Altug}},\ }\href {\doibase
  10.1126/science.aas9768} {\bibfield  {journal} {\bibinfo  {journal}
  {Science}\ }\textbf {\bibinfo {volume} {360}},\ \bibinfo {pages} {1105}
  (\bibinfo {year} {2018})}\BibitemShut {NoStop}%
\bibitem [{\citenamefont {Wang}\ \emph {et~al.}(2016)\citenamefont {Wang},
  \citenamefont {Ke}, \citenamefont {Xu}, \citenamefont {Zhan}, \citenamefont
  {Zheng}, \citenamefont {Wen}, \citenamefont {Yan}, \citenamefont {Liu},
  \citenamefont {Chen}, \citenamefont {She}, \citenamefont {Zhang},
  \citenamefont {Liu}, \citenamefont {Chen},\ and\ \citenamefont
  {Deng}}]{Strong_coupling1}%
  \BibitemOpen
  \bibfield  {author} {\bibinfo {author} {\bibfnamefont {H.}~\bibnamefont
  {Wang}}, \bibinfo {author} {\bibfnamefont {Y.}~\bibnamefont {Ke}}, \bibinfo
  {author} {\bibfnamefont {N.}~\bibnamefont {Xu}}, \bibinfo {author}
  {\bibfnamefont {R.}~\bibnamefont {Zhan}}, \bibinfo {author} {\bibfnamefont
  {Z.}~\bibnamefont {Zheng}}, \bibinfo {author} {\bibfnamefont
  {J.}~\bibnamefont {Wen}}, \bibinfo {author} {\bibfnamefont {J.}~\bibnamefont
  {Yan}}, \bibinfo {author} {\bibfnamefont {P.}~\bibnamefont {Liu}}, \bibinfo
  {author} {\bibfnamefont {J.}~\bibnamefont {Chen}}, \bibinfo {author}
  {\bibfnamefont {J.}~\bibnamefont {She}}, \bibinfo {author} {\bibfnamefont
  {Y.}~\bibnamefont {Zhang}}, \bibinfo {author} {\bibfnamefont
  {F.}~\bibnamefont {Liu}}, \bibinfo {author} {\bibfnamefont {H.}~\bibnamefont
  {Chen}}, \ and\ \bibinfo {author} {\bibfnamefont {S.}~\bibnamefont {Deng}},\
  }\href {\doibase 10.1021/acs.nanolett.6b02759} {\bibfield  {journal}
  {\bibinfo  {journal} {Nano Letters}\ }\textbf {\bibinfo {volume} {16}},\
  \bibinfo {pages} {6886} (\bibinfo {year} {2016})}\BibitemShut {NoStop}%
\bibitem [{\citenamefont {Ruan}\ \emph {et~al.}(2018)\citenamefont {Ruan},
  \citenamefont {Li}, \citenamefont {Yin}, \citenamefont {Cui}, \citenamefont
  {Wang},\ and\ \citenamefont {Lin}}]{Strong_coupling2}%
  \BibitemOpen
  \bibfield  {author} {\bibinfo {author} {\bibfnamefont {Q.}~\bibnamefont
  {Ruan}}, \bibinfo {author} {\bibfnamefont {N.}~\bibnamefont {Li}}, \bibinfo
  {author} {\bibfnamefont {H.}~\bibnamefont {Yin}}, \bibinfo {author}
  {\bibfnamefont {X.}~\bibnamefont {Cui}}, \bibinfo {author} {\bibfnamefont
  {J.}~\bibnamefont {Wang}}, \ and\ \bibinfo {author} {\bibfnamefont {H.-Q.}\
  \bibnamefont {Lin}},\ }\href {\doibase 10.1021/acsphotonics.8b00886}
  {\bibfield  {journal} {\bibinfo  {journal} {ACS Photonics}\ }\textbf
  {\bibinfo {volume} {5}},\ \bibinfo {pages} {3838} (\bibinfo {year}
  {2018})}\BibitemShut {NoStop}%
\bibitem [{\citenamefont {Tserkezis}\ \emph {et~al.}(2018)\citenamefont
  {Tserkezis}, \citenamefont {Gon\ifmmode~\mbox{\c{c}}\else \c{c}\fi{}alves},
  \citenamefont {Wolff}, \citenamefont {Todisco}, \citenamefont {Busch},\ and\
  \citenamefont {Mortensen}}]{Strong_coupling3}%
  \BibitemOpen
  \bibfield  {author} {\bibinfo {author} {\bibfnamefont {C.}~\bibnamefont
  {Tserkezis}}, \bibinfo {author} {\bibfnamefont {P.~A.~D.}\ \bibnamefont
  {Gon\ifmmode~\mbox{\c{c}}\else \c{c}\fi{}alves}}, \bibinfo {author}
  {\bibfnamefont {C.}~\bibnamefont {Wolff}}, \bibinfo {author} {\bibfnamefont
  {F.}~\bibnamefont {Todisco}}, \bibinfo {author} {\bibfnamefont
  {K.}~\bibnamefont {Busch}}, \ and\ \bibinfo {author} {\bibfnamefont {N.~A.}\
  \bibnamefont {Mortensen}},\ }\href {\doibase 10.1103/PhysRevB.98.155439}
  {\bibfield  {journal} {\bibinfo  {journal} {Phys. Rev. B}\ }\textbf {\bibinfo
  {volume} {98}},\ \bibinfo {pages} {155439} (\bibinfo {year}
  {2018})}\BibitemShut {NoStop}%
\bibitem [{\citenamefont {Todisco}\ \emph {et~al.}(2019)\citenamefont
  {Todisco}, \citenamefont {Malureanu}, \citenamefont {Wolff}, \citenamefont
  {Gonçalves}, \citenamefont {Roberts}, \citenamefont {Mortensen},\ and\
  \citenamefont {Tserkezis}}]{Strong_coupling4}%
  \BibitemOpen
  \bibfield  {author} {\bibinfo {author} {\bibfnamefont {F.}~\bibnamefont
  {Todisco}}, \bibinfo {author} {\bibfnamefont {R.}~\bibnamefont {Malureanu}},
  \bibinfo {author} {\bibfnamefont {C.}~\bibnamefont {Wolff}}, \bibinfo
  {author} {\bibfnamefont {P.~A.~D.}\ \bibnamefont {Gonçalves}}, \bibinfo
  {author} {\bibfnamefont {A.~S.}\ \bibnamefont {Roberts}}, \bibinfo {author}
  {\bibfnamefont {N.~A.}\ \bibnamefont {Mortensen}}, \ and\ \bibinfo {author}
  {\bibfnamefont {C.}~\bibnamefont {Tserkezis}},\ }\href@noop {} {\enquote
  {\bibinfo {title} {Magnetic and electric mie-exciton polaritons in silicon
  nanodisks},}\ } (\bibinfo {year} {2019})\BibitemShut {NoStop}%
\bibitem [{\citenamefont {Wang}\ \emph {et~al.}(2018)\citenamefont {Wang},
  \citenamefont {Titchener}, \citenamefont {Kruk}, \citenamefont {Xu},
  \citenamefont {Chung}, \citenamefont {Parry}, \citenamefont {Kravchenko},
  \citenamefont {Chen}, \citenamefont {Solntsev}, \citenamefont {Kivshar},
  \citenamefont {Neshev},\ and\ \citenamefont {Sukhorukov}}]{quant1}%
  \BibitemOpen
  \bibfield  {author} {\bibinfo {author} {\bibfnamefont {K.}~\bibnamefont
  {Wang}}, \bibinfo {author} {\bibfnamefont {J.~G.}\ \bibnamefont {Titchener}},
  \bibinfo {author} {\bibfnamefont {S.~S.}\ \bibnamefont {Kruk}}, \bibinfo
  {author} {\bibfnamefont {L.}~\bibnamefont {Xu}}, \bibinfo {author}
  {\bibfnamefont {H.-P.}\ \bibnamefont {Chung}}, \bibinfo {author}
  {\bibfnamefont {M.}~\bibnamefont {Parry}}, \bibinfo {author} {\bibfnamefont
  {I.~I.}\ \bibnamefont {Kravchenko}}, \bibinfo {author} {\bibfnamefont
  {Y.-H.}\ \bibnamefont {Chen}}, \bibinfo {author} {\bibfnamefont {A.~S.}\
  \bibnamefont {Solntsev}}, \bibinfo {author} {\bibfnamefont {Y.~S.}\
  \bibnamefont {Kivshar}}, \bibinfo {author} {\bibfnamefont {D.~N.}\
  \bibnamefont {Neshev}}, \ and\ \bibinfo {author} {\bibfnamefont {A.~A.}\
  \bibnamefont {Sukhorukov}},\ }\href {\doibase 10.1126/science.aat8196}
  {\bibfield  {journal} {\bibinfo  {journal} {Science}\ }\textbf {\bibinfo
  {volume} {361}},\ \bibinfo {pages} {1104} (\bibinfo {year}
  {2018})}\BibitemShut {NoStop}%
\bibitem [{\citenamefont {Stav}\ \emph {et~al.}(2018)\citenamefont {Stav},
  \citenamefont {Faerman}, \citenamefont {Maguid}, \citenamefont {Oren},
  \citenamefont {Kleiner}, \citenamefont {Hasman},\ and\ \citenamefont
  {Segev}}]{quant2}%
  \BibitemOpen
  \bibfield  {author} {\bibinfo {author} {\bibfnamefont {T.}~\bibnamefont
  {Stav}}, \bibinfo {author} {\bibfnamefont {A.}~\bibnamefont {Faerman}},
  \bibinfo {author} {\bibfnamefont {E.}~\bibnamefont {Maguid}}, \bibinfo
  {author} {\bibfnamefont {D.}~\bibnamefont {Oren}}, \bibinfo {author}
  {\bibfnamefont {V.}~\bibnamefont {Kleiner}}, \bibinfo {author} {\bibfnamefont
  {E.}~\bibnamefont {Hasman}}, \ and\ \bibinfo {author} {\bibfnamefont
  {M.}~\bibnamefont {Segev}},\ }\href {\doibase 10.1126/science.aat9042}
  {\bibfield  {journal} {\bibinfo  {journal} {Science}\ }\textbf {\bibinfo
  {volume} {361}},\ \bibinfo {pages} {1101} (\bibinfo {year}
  {2018})}\BibitemShut {NoStop}%
\bibitem [{\citenamefont {Kruk}\ \emph {et~al.}(2019)\citenamefont {Kruk},
  \citenamefont {Poddubny}, \citenamefont {Smirnova}, \citenamefont {Wang},
  \citenamefont {Slobozhanyuk}, \citenamefont {Shorokhov}, \citenamefont
  {Kravchenko}, \citenamefont {Luther-Davies},\ and\ \citenamefont
  {Kivshar}}]{quant3}%
  \BibitemOpen
  \bibfield  {author} {\bibinfo {author} {\bibfnamefont {S.}~\bibnamefont
  {Kruk}}, \bibinfo {author} {\bibfnamefont {A.}~\bibnamefont {Poddubny}},
  \bibinfo {author} {\bibfnamefont {D.}~\bibnamefont {Smirnova}}, \bibinfo
  {author} {\bibfnamefont {L.}~\bibnamefont {Wang}}, \bibinfo {author}
  {\bibfnamefont {A.}~\bibnamefont {Slobozhanyuk}}, \bibinfo {author}
  {\bibfnamefont {A.}~\bibnamefont {Shorokhov}}, \bibinfo {author}
  {\bibfnamefont {I.}~\bibnamefont {Kravchenko}}, \bibinfo {author}
  {\bibfnamefont {B.}~\bibnamefont {Luther-Davies}}, \ and\ \bibinfo {author}
  {\bibfnamefont {Y.}~\bibnamefont {Kivshar}},\ }\href {\doibase
  10.1038/s41565-018-0324-7} {\bibfield  {journal} {\bibinfo  {journal} {Nature
  Nanotechnology}\ }\textbf {\bibinfo {volume} {14}},\ \bibinfo {pages} {126}
  (\bibinfo {year} {2019})}\BibitemShut {NoStop}%
\bibitem [{\citenamefont {Jackson}(1999)}]{Jackson}%
  \BibitemOpen
  \bibfield  {author} {\bibinfo {author} {\bibfnamefont {J.~D.}\ \bibnamefont
  {Jackson}},\ }\href@noop {} {\emph {\bibinfo {title} {Classical
  Electrodynamics}}}\ (\bibinfo  {publisher} {John Wiley \& Sons, Inc.},\
  \bibinfo {address} {New York},\ \bibinfo {year} {1999})\BibitemShut {NoStop}%
\bibitem [{\citenamefont {Raab}\ and\ \citenamefont {de~Lange}(2005)}]{Raab}%
  \BibitemOpen
  \bibfield  {author} {\bibinfo {author} {\bibfnamefont {R.~E.}\ \bibnamefont
  {Raab}}\ and\ \bibinfo {author} {\bibfnamefont {O.~L.}\ \bibnamefont
  {de~Lange}},\ }\href@noop {} {\emph {\bibinfo {title} {Multipole Theory in
  Electromagnetism: Classical, Quantum, and Symmetry Aspects, with
  Applications}}}\ (\bibinfo  {publisher} {Oxford University Press on Demand},\
  \bibinfo {address} {Oxford},\ \bibinfo {year} {2005})\BibitemShut {NoStop}%
\bibitem [{\citenamefont {Radescu}\ and\ \citenamefont
  {Vaman}(2002)}]{Rodescu}%
  \BibitemOpen
  \bibfield  {author} {\bibinfo {author} {\bibfnamefont {E.~E.}\ \bibnamefont
  {Radescu}}\ and\ \bibinfo {author} {\bibfnamefont {G.}~\bibnamefont
  {Vaman}},\ }\href {\doibase 10.1103/PhysRevE.65.046609} {\bibfield  {journal}
  {\bibinfo  {journal} {Phys. Rev. E}\ }\textbf {\bibinfo {volume} {65}},\
  \bibinfo {pages} {046609} (\bibinfo {year} {2002})}\BibitemShut {NoStop}%
\bibitem [{\citenamefont {Dubovik}\ and\ \citenamefont
  {Tugushev}(1990)}]{Dubovik}%
  \BibitemOpen
  \bibfield  {author} {\bibinfo {author} {\bibfnamefont {V.}~\bibnamefont
  {Dubovik}}\ and\ \bibinfo {author} {\bibfnamefont {V.}~\bibnamefont
  {Tugushev}},\ }\href {\doibase https://doi.org/10.1016/0370-1573(90)90042-Z}
  {\bibfield  {journal} {\bibinfo  {journal} {Physics Reports}\ }\textbf
  {\bibinfo {volume} {187}},\ \bibinfo {pages} {145} (\bibinfo {year}
  {1990})}\BibitemShut {NoStop}%
\bibitem [{\citenamefont {Nemkov}\ \emph {et~al.}(2018)\citenamefont {Nemkov},
  \citenamefont {Basharin},\ and\ \citenamefont {Fedotov}}]{Basharin}%
  \BibitemOpen
  \bibfield  {author} {\bibinfo {author} {\bibfnamefont {N.~A.}\ \bibnamefont
  {Nemkov}}, \bibinfo {author} {\bibfnamefont {A.~A.}\ \bibnamefont
  {Basharin}}, \ and\ \bibinfo {author} {\bibfnamefont {V.~A.}\ \bibnamefont
  {Fedotov}},\ }\href {\doibase 10.1103/PhysRevA.98.023858} {\bibfield
  {journal} {\bibinfo  {journal} {Phys. Rev. A}\ }\textbf {\bibinfo {volume}
  {98}},\ \bibinfo {pages} {023858} (\bibinfo {year} {2018})}\BibitemShut
  {NoStop}%
\bibitem [{\citenamefont {Zografopoulos}\ \emph {et~al.}(2019)\citenamefont
  {Zografopoulos}, \citenamefont {Algorri}, \citenamefont {Ferraro},
  \citenamefont {Garc{\'i}a-C{\'a}imara}, \citenamefont {S{\'a}inchez-Pena},\
  and\ \citenamefont {Beccherelli}}]{Beccherelli}%
  \BibitemOpen
  \bibfield  {author} {\bibinfo {author} {\bibfnamefont {D.~C.}\ \bibnamefont
  {Zografopoulos}}, \bibinfo {author} {\bibfnamefont {J.~F.}\ \bibnamefont
  {Algorri}}, \bibinfo {author} {\bibfnamefont {A.}~\bibnamefont {Ferraro}},
  \bibinfo {author} {\bibfnamefont {B.}~\bibnamefont {Garc{\'i}a-C{\'a}imara}},
  \bibinfo {author} {\bibfnamefont {J.~M.}\ \bibnamefont {S{\'a}inchez-Pena}},
  \ and\ \bibinfo {author} {\bibfnamefont {R.}~\bibnamefont {Beccherelli}},\
  }\href {\doibase 10.1038/s41598-019-44093-7} {\bibfield  {journal} {\bibinfo
  {journal} {Scientific Reports}\ }\textbf {\bibinfo {volume} {9}},\ \bibinfo
  {pages} {7544} (\bibinfo {year} {2019})}\BibitemShut {NoStop}%
\bibitem [{\citenamefont {Gurvitz}\ \emph {et~al.}(2019)\citenamefont
  {Gurvitz}, \citenamefont {Ladutenko}, \citenamefont {Dergachev},
  \citenamefont {Evlyukhin}, \citenamefont {Miroshnichenko},\ and\
  \citenamefont {Shalin}}]{Gurvitz2019}%
  \BibitemOpen
  \bibfield  {author} {\bibinfo {author} {\bibfnamefont {E.~A.}\ \bibnamefont
  {Gurvitz}}, \bibinfo {author} {\bibfnamefont {K.~S.}\ \bibnamefont
  {Ladutenko}}, \bibinfo {author} {\bibfnamefont {P.~A.}\ \bibnamefont
  {Dergachev}}, \bibinfo {author} {\bibfnamefont {A.~B.}\ \bibnamefont
  {Evlyukhin}}, \bibinfo {author} {\bibfnamefont {A.~E.}\ \bibnamefont
  {Miroshnichenko}}, \ and\ \bibinfo {author} {\bibfnamefont {A.~S.}\
  \bibnamefont {Shalin}},\ }\href {\doibase 10.1002/lpor.201800266} {\bibfield
  {journal} {\bibinfo  {journal} {Laser \& Photonics Reviews}\ }\textbf
  {\bibinfo {volume} {13}},\ \bibinfo {pages} {1800266} (\bibinfo {year}
  {2019})}\BibitemShut {NoStop}%
\bibitem [{\citenamefont {Grahn}\ \emph {et~al.}(2012)\citenamefont {Grahn},
  \citenamefont {Shevchenko},\ and\ \citenamefont {Kaivola}}]{Grahn}%
  \BibitemOpen
  \bibfield  {author} {\bibinfo {author} {\bibfnamefont {P.}~\bibnamefont
  {Grahn}}, \bibinfo {author} {\bibfnamefont {A.}~\bibnamefont {Shevchenko}}, \
  and\ \bibinfo {author} {\bibfnamefont {M.}~\bibnamefont {Kaivola}},\ }\href
  {\doibase 10.1088/1367-2630/14/9/093033} {\bibfield  {journal} {\bibinfo
  {journal} {New Journal of Physics}\ }\textbf {\bibinfo {volume} {14}},\
  \bibinfo {pages} {093033} (\bibinfo {year} {2012})}\BibitemShut {NoStop}%
\bibitem [{\citenamefont {Evlyukhin}\ \emph {et~al.}(2016)\citenamefont
  {Evlyukhin}, \citenamefont {Fischer}, \citenamefont {Reinhardt},\ and\
  \citenamefont {Chichkov}}]{Evlyukhin2016}%
  \BibitemOpen
  \bibfield  {author} {\bibinfo {author} {\bibfnamefont {A.~B.}\ \bibnamefont
  {Evlyukhin}}, \bibinfo {author} {\bibfnamefont {T.}~\bibnamefont {Fischer}},
  \bibinfo {author} {\bibfnamefont {C.}~\bibnamefont {Reinhardt}}, \ and\
  \bibinfo {author} {\bibfnamefont {B.~N.}\ \bibnamefont {Chichkov}},\ }\href
  {\doibase 10.1103/PhysRevB.94.205434} {\bibfield  {journal} {\bibinfo
  {journal} {Phys. Rev. B}\ }\textbf {\bibinfo {volume} {94}},\ \bibinfo
  {pages} {205434} (\bibinfo {year} {2016})}\BibitemShut {NoStop}%
\bibitem [{\citenamefont {Alaee}\ \emph {et~al.}(2018)\citenamefont {Alaee},
  \citenamefont {Rockstuhl},\ and\ \citenamefont
  {Fernandez-Corbaton}}]{Rockschtul2018}%
  \BibitemOpen
  \bibfield  {author} {\bibinfo {author} {\bibfnamefont {R.}~\bibnamefont
  {Alaee}}, \bibinfo {author} {\bibfnamefont {C.}~\bibnamefont {Rockstuhl}}, \
  and\ \bibinfo {author} {\bibfnamefont {I.}~\bibnamefont
  {Fernandez-Corbaton}},\ }\href {\doibase
  https://doi.org/10.1016/j.optcom.2017.08.064} {\bibfield  {journal} {\bibinfo
   {journal} {Optics Communications}\ }\textbf {\bibinfo {volume} {407}},\
  \bibinfo {pages} {17 } (\bibinfo {year} {2018})}\BibitemShut {NoStop}%
\bibitem [{\citenamefont {Alaee}\ \emph {et~al.}(2019)\citenamefont {Alaee},
  \citenamefont {Rockstuhl},\ and\ \citenamefont
  {Fernandez-Corbaton}}]{Rockschtul2019}%
  \BibitemOpen
  \bibfield  {author} {\bibinfo {author} {\bibfnamefont {R.}~\bibnamefont
  {Alaee}}, \bibinfo {author} {\bibfnamefont {C.}~\bibnamefont {Rockstuhl}}, \
  and\ \bibinfo {author} {\bibfnamefont {I.}~\bibnamefont
  {Fernandez-Corbaton}},\ }\href {\doibase 10.1002/adom.201800783} {\bibfield
  {journal} {\bibinfo  {journal} {Advanced Optical Materials}\ }\textbf
  {\bibinfo {volume} {7}},\ \bibinfo {pages} {1800783} (\bibinfo {year}
  {2019})}\BibitemShut {NoStop}%
\bibitem [{\citenamefont {Evlyukhin}\ and\ \citenamefont
  {Chichkov}(2019)}]{Evlyukhin2019}%
  \BibitemOpen
  \bibfield  {author} {\bibinfo {author} {\bibfnamefont {A.~B.}\ \bibnamefont
  {Evlyukhin}}\ and\ \bibinfo {author} {\bibfnamefont {B.~N.}\ \bibnamefont
  {Chichkov}},\ }\href {\doibase 10.1103/PhysRevB.100.125415} {\bibfield
  {journal} {\bibinfo  {journal} {Phys. Rev. B}\ }\textbf {\bibinfo {volume}
  {100}},\ \bibinfo {pages} {125415} (\bibinfo {year} {2019})}\BibitemShut
  {NoStop}%
\bibitem [{\citenamefont {Mie}(1908)}]{Mie}%
  \BibitemOpen
  \bibfield  {author} {\bibinfo {author} {\bibfnamefont {G.}~\bibnamefont
  {Mie}},\ }\href {\doibase 10.1002/andp.19083300302} {\bibfield  {journal}
  {\bibinfo  {journal} {Annalen der Physik}\ }\textbf {\bibinfo {volume}
  {330}},\ \bibinfo {pages} {377} (\bibinfo {year} {1908})}\BibitemShut
  {NoStop}%
\bibitem [{\citenamefont {Bohren}\ and\ \citenamefont
  {Huffman}(1998)}]{Bohren}%
  \BibitemOpen
  \bibfield  {author} {\bibinfo {author} {\bibfnamefont {C.~F.}\ \bibnamefont
  {Bohren}}\ and\ \bibinfo {author} {\bibfnamefont {D.~R.}\ \bibnamefont
  {Huffman}},\ }\href@noop {} {\emph {\bibinfo {title} {Absorption and
  Scattering of Light by Small Particles}}}\ (\bibinfo  {publisher} {John Wiley
  \& Sons, Inc.},\ \bibinfo {address} {New York},\ \bibinfo {year}
  {1998})\BibitemShut {NoStop}%
\bibitem [{\citenamefont {Zenin}\ \emph {et~al.}(2017)\citenamefont {Zenin},
  \citenamefont {Evlyukhin}, \citenamefont {Novikov}, \citenamefont {Yang},
  \citenamefont {Malureanu}, \citenamefont {Lavrinenko}, \citenamefont
  {Chichkov},\ and\ \citenamefont {Bozhevolnyi}}]{anapole}%
  \BibitemOpen
  \bibfield  {author} {\bibinfo {author} {\bibfnamefont {V.~A.}\ \bibnamefont
  {Zenin}}, \bibinfo {author} {\bibfnamefont {A.~B.}\ \bibnamefont
  {Evlyukhin}}, \bibinfo {author} {\bibfnamefont {S.~M.}\ \bibnamefont
  {Novikov}}, \bibinfo {author} {\bibfnamefont {Y.}~\bibnamefont {Yang}},
  \bibinfo {author} {\bibfnamefont {R.}~\bibnamefont {Malureanu}}, \bibinfo
  {author} {\bibfnamefont {A.~V.}\ \bibnamefont {Lavrinenko}}, \bibinfo
  {author} {\bibfnamefont {B.~N.}\ \bibnamefont {Chichkov}}, \ and\ \bibinfo
  {author} {\bibfnamefont {S.~I.}\ \bibnamefont {Bozhevolnyi}},\ }\href
  {\doibase 10.1021/acs.nanolett.7b04200} {\bibfield  {journal} {\bibinfo
  {journal} {Nano Letters}\ }\textbf {\bibinfo {volume} {17}},\ \bibinfo
  {pages} {7152} (\bibinfo {year} {2017})}\BibitemShut {NoStop}%
\bibitem [{\citenamefont {Yang}\ \emph {et~al.}(2018)\citenamefont {Yang},
  \citenamefont {Zenin},\ and\ \citenamefont {Bozhevolnyi}}]{slit}%
  \BibitemOpen
  \bibfield  {author} {\bibinfo {author} {\bibfnamefont {Y.}~\bibnamefont
  {Yang}}, \bibinfo {author} {\bibfnamefont {V.~A.}\ \bibnamefont {Zenin}}, \
  and\ \bibinfo {author} {\bibfnamefont {S.~I.}\ \bibnamefont {Bozhevolnyi}},\
  }\href {\doibase 10.1021/acsphotonics.7b01440} {\bibfield  {journal}
  {\bibinfo  {journal} {ACS Photonics}\ }\textbf {\bibinfo {volume} {5}},\
  \bibinfo {pages} {1960} (\bibinfo {year} {2018})}\BibitemShut {NoStop}%
\bibitem [{\citenamefont {Burrows}\ and\ \citenamefont
  {Barnes}(2010)}]{Barnes2010}%
  \BibitemOpen
  \bibfield  {author} {\bibinfo {author} {\bibfnamefont {C.~P.}\ \bibnamefont
  {Burrows}}\ and\ \bibinfo {author} {\bibfnamefont {W.~L.}\ \bibnamefont
  {Barnes}},\ }\href {\doibase 10.1364/OE.18.003187} {\bibfield  {journal}
  {\bibinfo  {journal} {Opt. Express}\ }\textbf {\bibinfo {volume} {18}},\
  \bibinfo {pages} {3187} (\bibinfo {year} {2010})}\BibitemShut {NoStop}%
\bibitem [{\citenamefont {Born}\ and\ \citenamefont {Wolf}(1999)}]{Wolf}%
  \BibitemOpen
  \bibfield  {author} {\bibinfo {author} {\bibfnamefont {M.}~\bibnamefont
  {Born}}\ and\ \bibinfo {author} {\bibfnamefont {E.}~\bibnamefont {Wolf}},\
  }\href {\doibase 10.1017/CBO9781139644181} {\emph {\bibinfo {title}
  {Principles of Optics}}},\ \bibinfo {edition} {7th}\ ed.\ (\bibinfo
  {publisher} {Cambridge University Press},\ \bibinfo {address} {Cambridge},\
  \bibinfo {year} {1999})\BibitemShut {NoStop}%
\end{thebibliography}%

\pagebreak
\widetext
\section*{Supporting information}
\setcounter{equation}{0}
\setcounter{figure}{0}
\setcounter{table}{0}
\setcounter{page}{1}
\makeatletter
\renewcommand{\thepage}{s\arabic{page}}
\renewcommand{\theequation}{S\arabic{equation}}
\renewcommand{\thefigure}{S\arabic{figure}}

\textbf{Supplementary Note 1: analytical expressions for multipole scattering diagrams}

The angular dependence of the scattered field in the far-field zone for a perfect sphere in an even dielectric environment is following \cite{Barnes2010}:
\begin{equation}
\label{EqS1}
\left\{ \begin{aligned}
  & {{E}_{\theta }}\left( \theta ,\varphi  \right)\approx {{E}_{0}}\frac{{{e}^{ikr}}}{-ikr}\cos \varphi \sum\limits_{n}{\frac{2n+1}{n\left( n+1 \right)}}\left( {{a}_{n}}\frac{dP_{n}^{1}}{d\theta }+{{b}_{n}}\frac{P_{n}^{1}}{\sin \theta } \right), \\ 
 & {{E}_{\varphi }}\left( \theta ,\varphi  \right)\approx -{{E}_{0}}\frac{{{e}^{ikr}}}{-ikr}\sin \varphi \sum\limits_{n}{\frac{2n+1}{n\left( n+1 \right)}}\left( {{a}_{n}}\frac{P_{n}^{1}}{\sin \theta }+{{b}_{n}}\frac{dP_{n}^{1}}{d\theta } \right), \\ 
\end{aligned} \right.
\end{equation}                        
where $\theta$ and $\phi$ are the polar and azimuthal scattering angles, respectively, $E_0$ is the strength of the incident field, $k$ is the wavevector in the surrounding environment, and $n$ is the order of the multipole (e.g., $n$ = 1 defines dipole, $n$ = 2 defines quadrupole, etc.). $a_n$ ($b_n$) are the complex Mie coefficients, corresponding to the electric (magnetic) multipoles. $P_{n}^{1}$ represents the set of associated Legendre polynomials of order 1, which are convenient to rewrite using following angle-dependent functions \cite{Bohren}: 
\begin{equation*}
{{\pi }_{n}}\left( \theta  \right)\equiv \frac{P_{n}^{1}}{\sin \theta },\text{\ \ \ \ \ \ \ }{{\tau }_{n}}\left( \theta  \right)\equiv \frac{dP_{n}^{1}}{d\theta }
\end{equation*}
They can be computed iteratively using following relations:
\begin{equation*}
\left\{ \begin{aligned}
  & {{\pi }_{n}}=\frac{2n-1}{n-1}\cos \theta {{\pi }_{n-1}}-\frac{n}{n-1}{{\pi }_{n-2}}, \\ 
 & {{\tau }_{n}}=n\cos \theta {{\pi }_{n}}-\left( n+1 \right){{\pi }_{n-1}}, \\ 
 & {{\pi }_{0}}=0,\text{      }{{\pi }_{1}}=1. \\ 
\end{aligned} \right.
\end{equation*}
The scattering cross-section into a unit solid angle is proportional to the square of the $E$-field:
\begin{equation}
\label{EqS2}
\sigma \left( \theta ,\varphi  \right)\propto {{\left| {{E}_{\theta }}\left( \theta ,\varphi  \right) \right|}^{2}}+{{\left| {{E}_{\varphi }}\left( \theta ,\varphi  \right) \right|}^{2}},
\end{equation}                                               
therefore the scattering diagram from the given multipole of order n can be written as following:
\begin{equation*}
\left\{ \begin{aligned}
  & {{\sigma }_{\text{En}}}\left( \theta ,\varphi  \right)\propto {{\cos }^{2}}\varphi {{\tau }_{n}}^{2}+{{\sin }^{2}}\varphi {{\pi }_{n}}^{2}, \\ 
 & {{\sigma }_{\text{Mn}}}\left( \theta ,\varphi  \right)\propto {{\sin }^{2}}\varphi {{\tau }_{n}}^{2}+{{\cos }^{2}}\varphi {{\pi }_{n}}^{2}={{\sigma }_{\text{En}}}\left( \theta ,\tfrac{1}{2}\pi -\varphi  \right). \\ 
\end{aligned} \right.
\end{equation*}
One can see that these angle-dependent functions define the dependence of the scattering cross-section on the polar angle $\theta$ within cross-sections at $\phi = 0$ at $\phi = \pi/2$. The table below collects explicit expressions for angle-dependent functions, 3D illustration of the scattering diagram, and its corresponding sections at $\phi = 0$ at $\phi = \pi/2$ for the first four orders of multipoles:

\centerline{\includegraphics{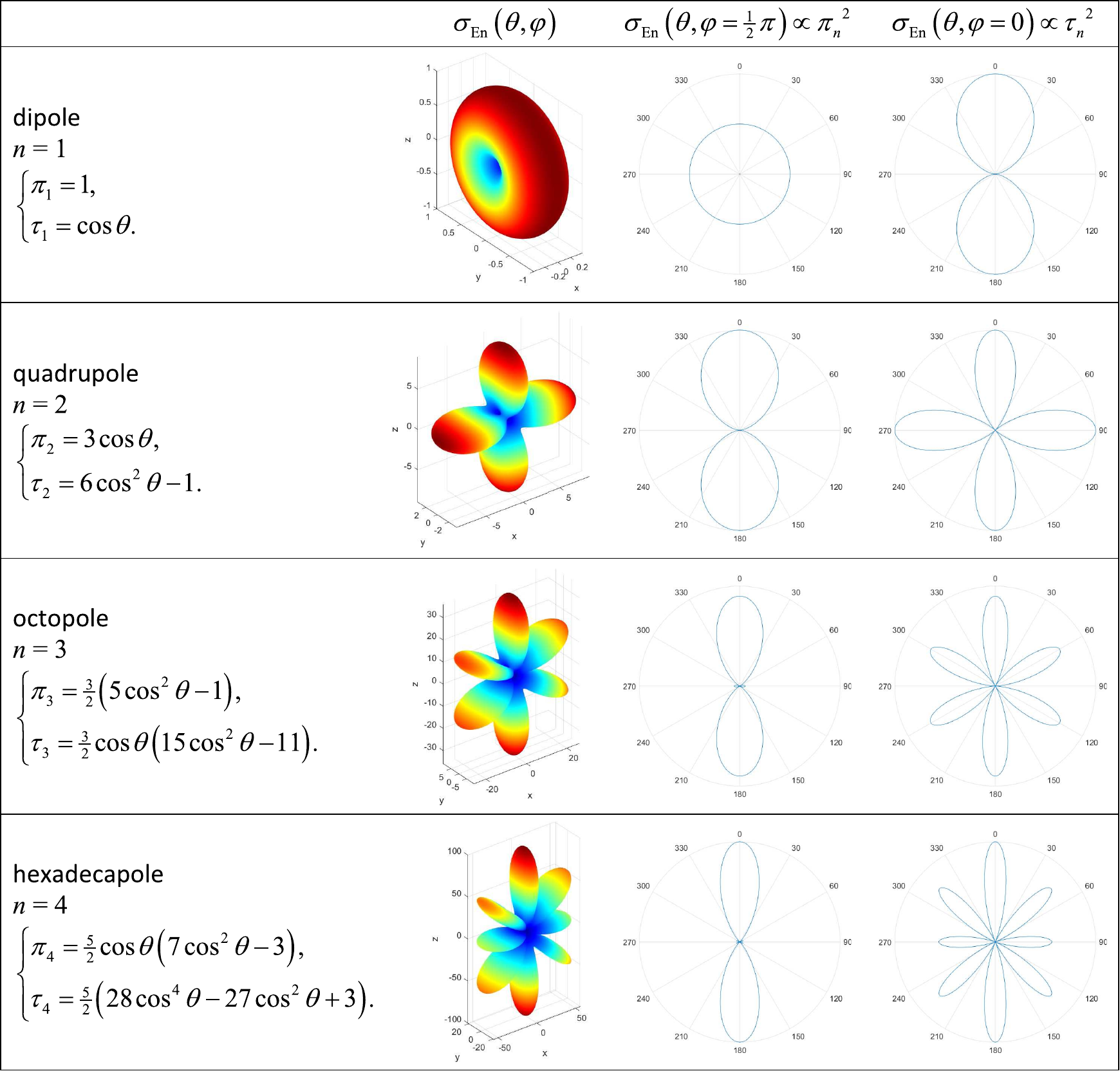}}
Table S1: scattering diagrams of the first four multipole orders and their correspondent angle-dependent functions.
\\

One can see that the principal scattering lobes for any multipole order $n > 1$ are defined by the angle-dependent function $\tau_{n}$. Their position $\theta_{\rm max}$ can be easily found by taking the derivative of  $\tau_{n}$:
\begin{equation*}
{{\left. \frac{d{{\tau }_{n}}}{d\theta } \right|}_{\theta =\theta \max }}=0
\end{equation*}
Angles of non-trivial scattering lobes for octopole ($n$ = 3) and hexadecapole ($n$ = 4) are thus:
\begin{equation}
\label{EqS3}
\left\{ \begin{aligned}
  & {{\theta }_{\text{O,}\max }}=\arccos \left( \frac{1}{2}\sqrt{\frac{44}{45}} \right)\approx \frac{\pi }{3}, \\ 
 & {{\theta }_{\text{H,}\max }}=\arccos \left( \frac{1}{\sqrt{2}}\sqrt{\frac{27}{28}} \right)\approx \frac{\pi }{4}. \\ 
\end{aligned} \right.
\end{equation}
Regarding the number of all scattering lobes, one can actually find two additional tiny scattering lobes for electric (magnetic) octopole and four for electric (magnetic) hexadecapole at $\phi = \pi/2$ ($\phi = 0$) section, defined by the other angle-dependent function $\pi_n$. Thus the number of scattering lobes would depend on their definition, and we believe it has no fundamental direct correlation with the multipole order.
The total scattering cross-section can be calculated as following \cite{Bohren}:
\begin{equation*}
{{\sigma }_{\text{total}}}=\frac{2\pi }{{{k}^{2}}}\sum\limits_{n}{\left( 2n+1 \right)}\left( {{\left| {{a}_{n}} \right|}^{2}}+{{\left| {{b}_{n}} \right|}^{2}} \right),
\end{equation*}                                                    
from which one can determine individual contributions from each multipole.

Assume that only magnetic dipole (MD) and electric octopole (EO) are present. The asymmetry between forward ($\theta = 0$) and backward scattering ($\theta = \pi$) can be calculated using Eqs.~\ref{EqS1} and \ref{EqS2}:
\begin{equation*}
\frac{{{\sigma }_{\text{forward}}}}{{{\sigma }_{\text{backward}}}}=\frac{\sigma \left( \theta =0,\varphi  \right)}{\sigma \left( \theta =\pi ,\varphi  \right)}={{\left| \frac{\tfrac{7}{2}{{a}_{\text{EO}}}+\tfrac{3}{2}{{b}_{\text{MD}}}}{\tfrac{7}{2}{{a}_{\text{EO}}}-\tfrac{3}{2}{{b}_{\text{MD}}}} \right|}^{2}}={{\left| \frac{1+\tfrac{3}{7}{{b}_{\text{MD}}}/{{a}_{\text{EO}}}}{1-\tfrac{3}{7}{{b}_{\text{MD}}}/{{a}_{\text{EO}}}} \right|}^{2}}
\end{equation*}
From Eq.~\ref{EqS3} it follows that
\begin{equation*}
\frac{\left| {{b}_{\text{MD}}} \right|}{\left| {{a}_{\text{EO}}} \right|}=\sqrt{\frac{7}{3}\frac{{{\sigma }_{\text{MD}}}}{{{\sigma }_{\text{EO}}}}},
\end{equation*}
If backward scattering cannot be completely canceled, then the maximum asymmetry in forward/backward scattering is:
\begin{equation}
\label{EqS4}
\max \left\{ \frac{{{\sigma }_{\text{forward}}}}{{{\sigma }_{\text{backward}}}} \right\}={{\left| \frac{1+\sqrt{\tfrac{3}{7}{{{\sigma }_{\text{MD}}}}/{{{\sigma }_{\text{EO}}}}\;}}{1-\sqrt{\tfrac{3}{7}{{{\sigma }_{\text{MD}}}}/{{{\sigma }_{\text{EO}}}}\;}} \right|}^{2}}
\end{equation}                                                  

\clearpage

\begin{figure}
\centering\includegraphics{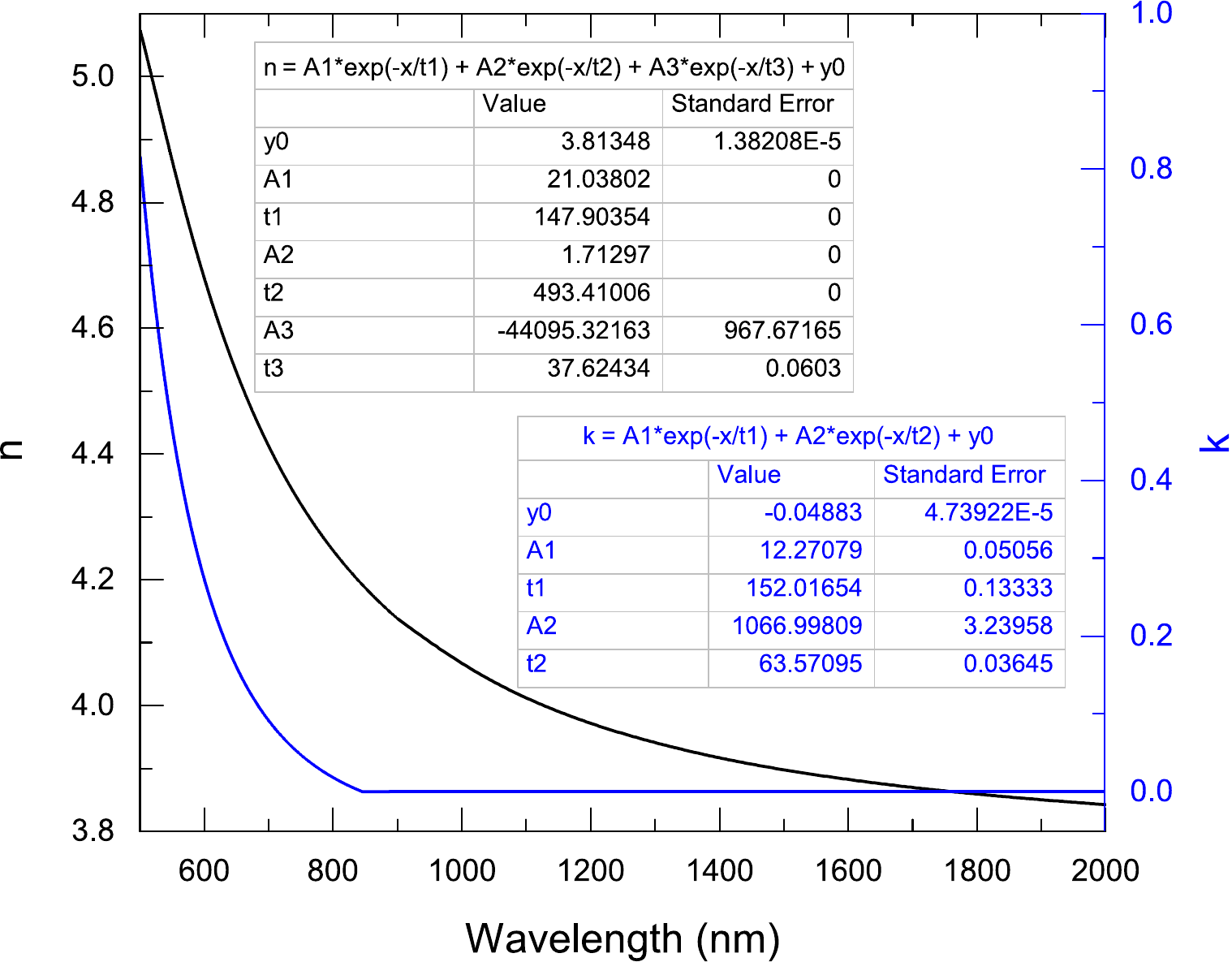}
  \caption{
	Refractive index of the fabricated amorphous Si, measured by reflectometry, along with fitting functions.
	}
	\label{S1}
\end{figure}

\begin{figure}
\centering\includegraphics{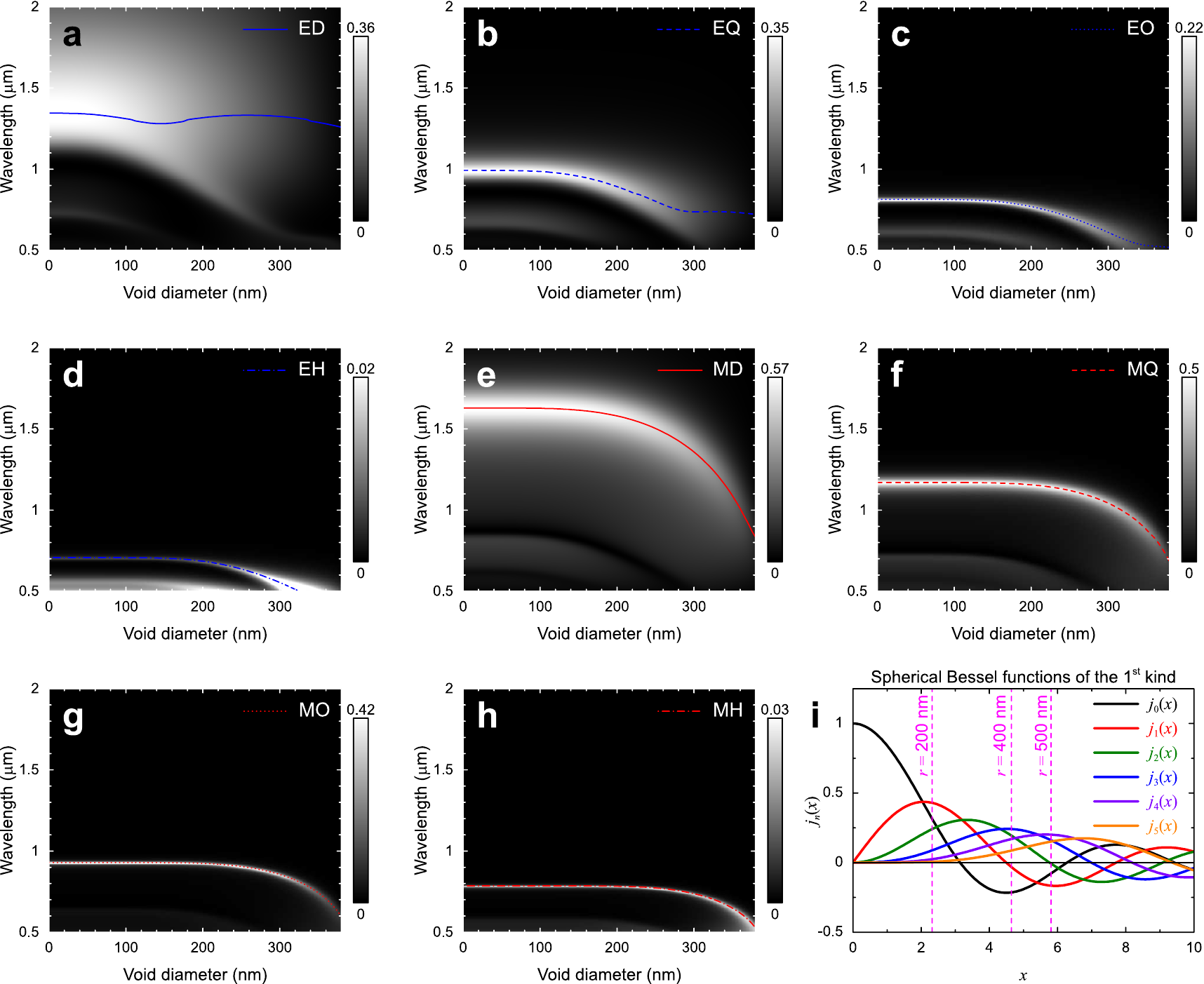}
  \caption{
	Electric (a-d) and magnetic multipole (e-h) contributions to the total scattering cross-section as a function of the free-space wavelength and void diameter for Si sphere/shell with outer diameter of 400 nm. D stands for dipole, Q – quadrupole, O – octopole, and H – hexadecapole. Resonances (position of local maximum) for electric (blue) and magnetic multipoles (red) are shown with lines. (i) Spherical Bessel functions of the first kind. Magenta lines are drawn at $x = k_{800}r = 2\pi rn/\lambda_{800}$, where $k_{800}$ is a wavenumber, $n = 1.48$ is the refractive index of the environment, and $\lambda_{800}$ = 800 nm is a free-space wavelength. 
	}
  \label{S2}
\end{figure}

\begin{figure}
\centering\includegraphics{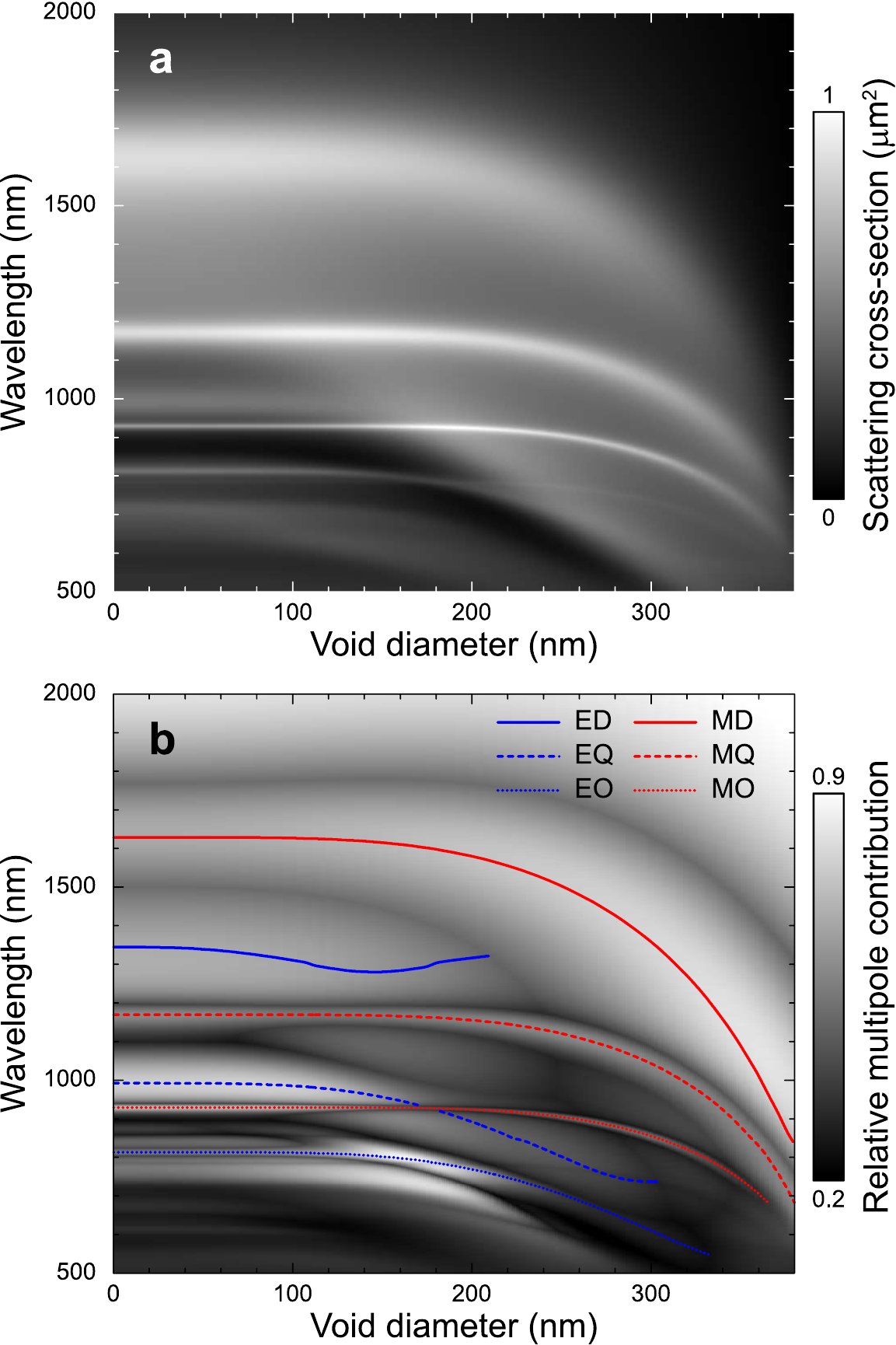}
  \caption{
	(a) Total scattering cross-section as a function of the free-space wavelength and void diameter for Si sphere/shell with outer diameter of 400 nm. (b) Relative contribution from a dominant multipole to the total scattering cross-section for Si sphere/shell particle in $n = 1.48$ environment. Resonances for electric (blue) and magnetic multipoles (red) are shown with lines in the domain where their contribution is dominant. 
	}
  \label{S3}
\end{figure}

\begin{figure}
\centering\includegraphics{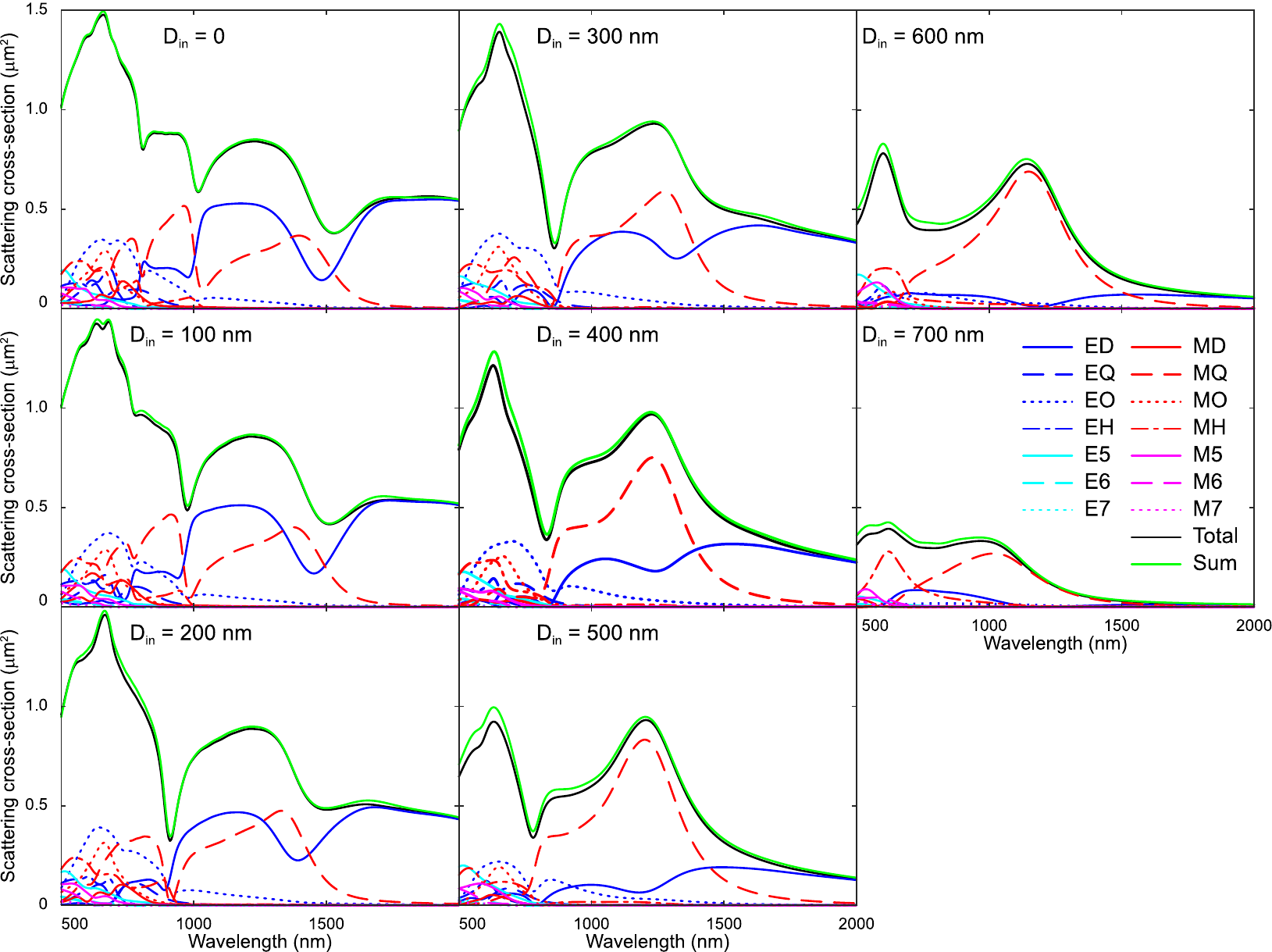}
  \caption{
	Total scattering cross-section (black) and individual contributions from electric (blue) and magnetic multipoles (red) for the silicon ring with the outer diameter of 800 nm and varied internal diameter. The sum of contributions from all considered multipoles is plotted with a green line. The ring thickness is 80 nm, and the refractive index of surrounding is $n = 1.48$. 
	}
  \label{S4}
\end{figure}

\begin{figure}
\centering\includegraphics{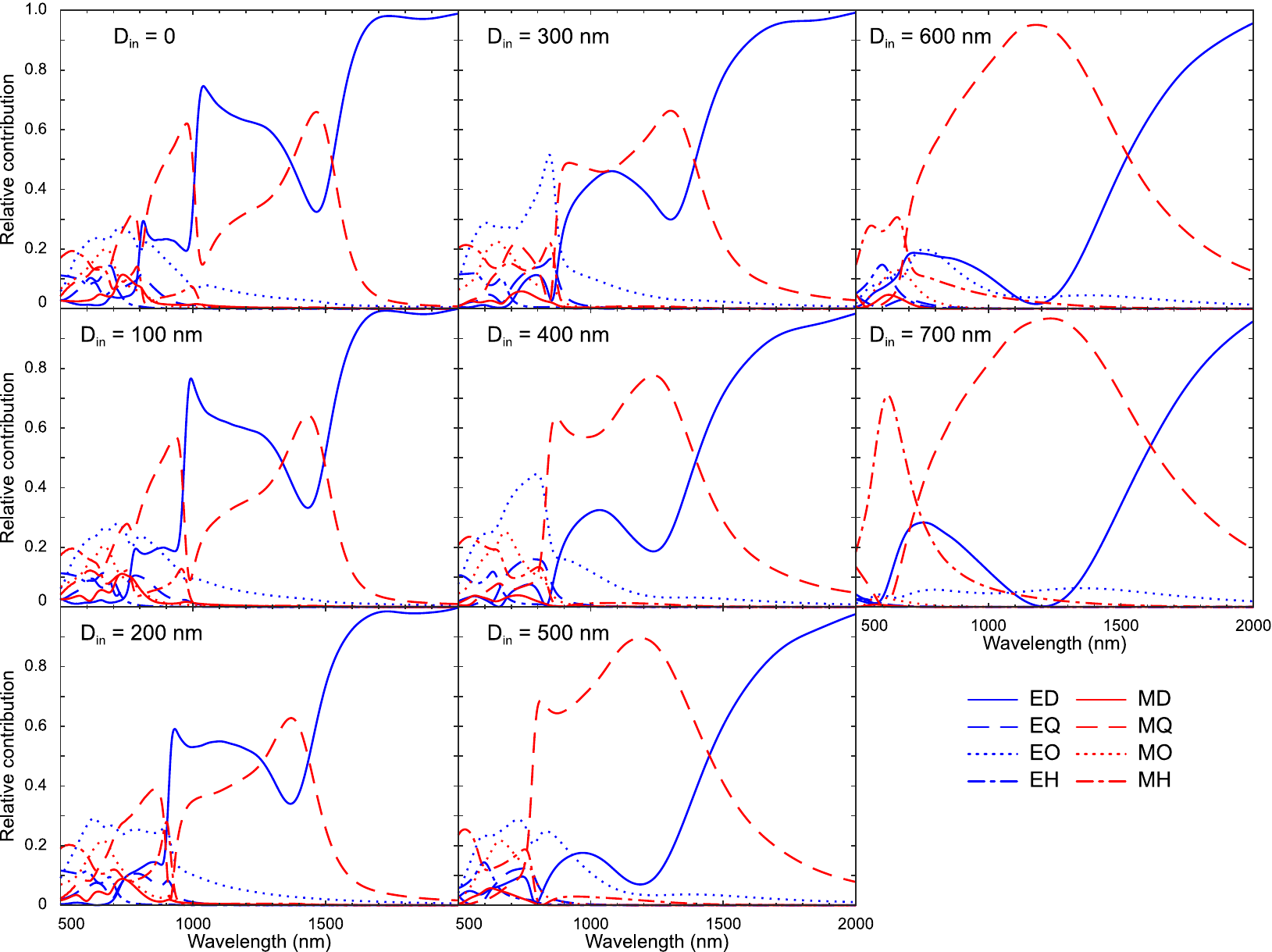}
  \caption{
	Relative contributions from electric (blue) and magnetic multipoles (red) to the total scattering cross-section for the silicon ring with the outer diameter of 800 nm and varied internal diameter. The ring thickness is 80 nm, and the refractive index of surrounding is $n = 1.48$. 
	}
  \label{S5}
\end{figure}

\begin{figure}
\centering\includegraphics{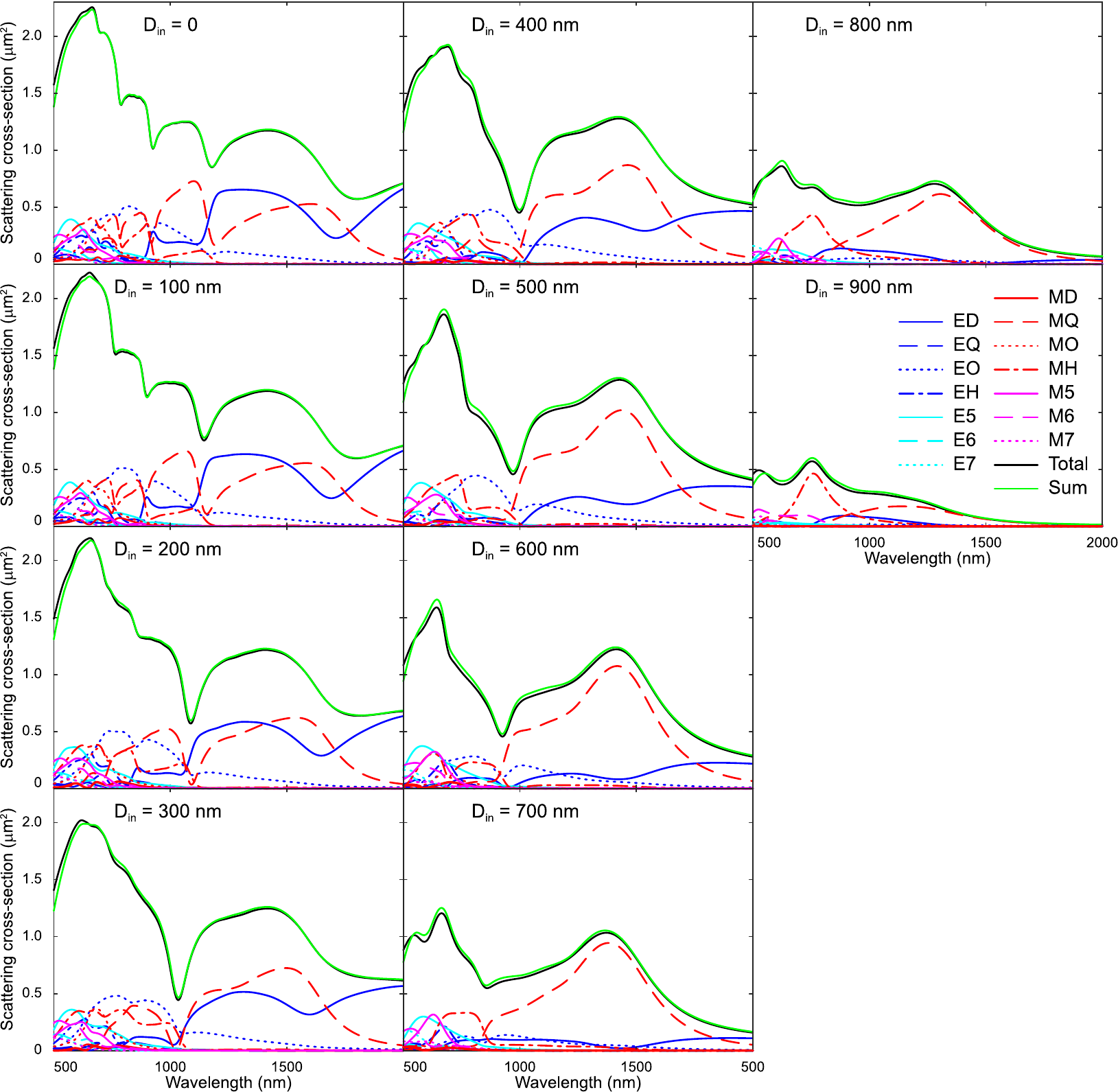}
  \caption{
	Total scattering cross-section (black) and individual contributions from electric (blue) and magnetic multipoles (red) for the silicon ring with the outer diameter of 1000 nm and varied internal diameter. The sum of contributions from all considered multipoles is plotted with a green line. The ring thickness is 80 nm, and the refractive index of surrounding is $n = 1.48$. 
	}
  \label{S6}
\end{figure}

\begin{figure}
\centering\includegraphics{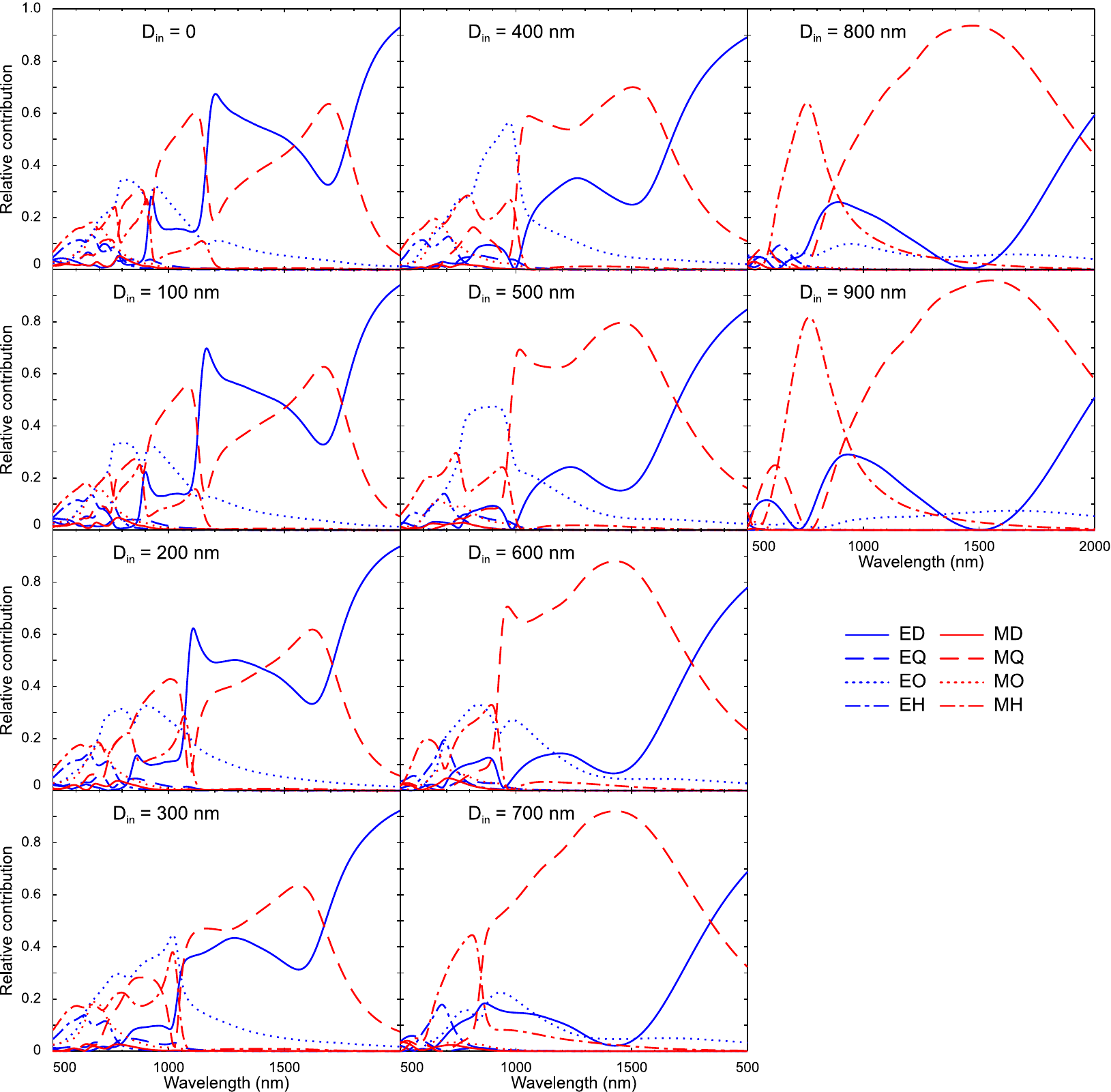}
  \caption{
	Relative contributions from electric (blue) and magnetic multipoles (red) to the total scattering cross-section for the silicon ring with the outer diameter of 1000 nm and varied internal diameter. The ring thickness is 80 nm, and the refractive index of surrounding is $n = 1.48$. 
	}
  \label{S7}
\end{figure}

\begin{figure}
\centering\includegraphics{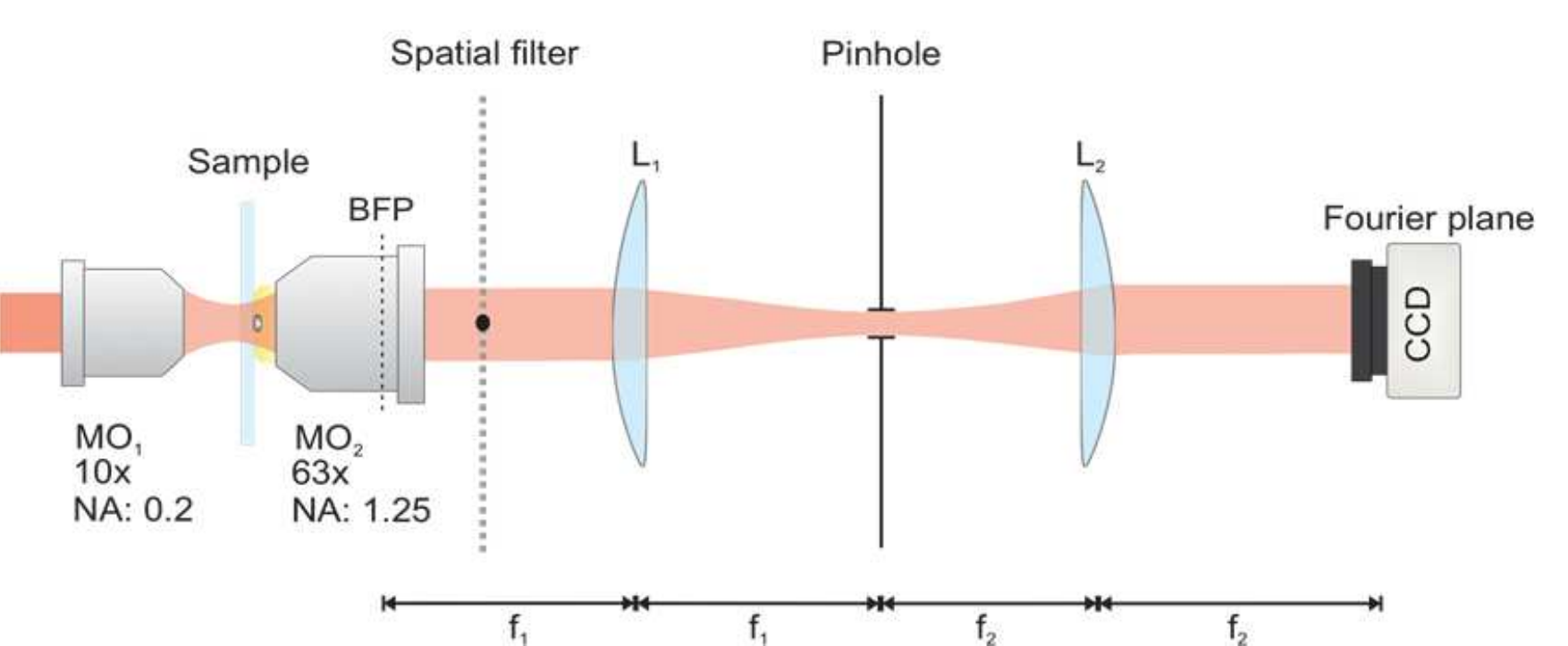}
  \caption{
	Schematic diagram of the experimental setup used to measure the scattering diagrams. The sample was illuminated using a linearly polarized Ti:Sapphire laser, tuned at a wavelength of $\sim$800 nm. The laser beam was weakly focused onto the sample using a 10$\times$ objective of numerical aperture NA = 0.20. The full-width-at-half-maximum (FHWM) of the focused beam spot was $\sim$5 $\mu$m. The scattered light was collected using a 63$\times$ oil-immersion collection objective, with a NA = 1.25. The structures on the sample were positioned facing the oil-immersion objective, embedded in the index-matching oil ($n = 1.518$). An imaging system, which consists of two lenses and two spatial filters, was used to image the back focal plane (BFP) of the collection objective with a charge-coupled device (CCD) camera. The BFP and its image are also referred to here as the Fourier plane, because it shows angular distribution of the scattering (i.e., scattering diagram). The two spatial filters have the following important functionalities. The first filter is a micrometric metallic ball (diameter $\sim$300 $\mu$m), glued on a glass substrate, and is used to stop the directly-transmitted light from reaching the CCD camera to avoid saturation. It is a Fourier-plane filter, and ideally it should be placed at the BFP inside the collection objective. Nevertheless, since the objective collimates the scattered light, the filter can be placed at the rear aperture of the objective and still produce the same filtering effect. The second filter is a pinhole, positioned at the image plane (focal point of the first lens), and is used to stop all the unwanted scattering from the surroundings of the nanostructure (i.e., from impurities in the glass substrate and oil). The images are captured with the CCD camera, located at the focal distance of the second lens, which matches with the Fourier plane.
	}
  \label{S8}
\end{figure}

\begin{figure}
\centering\includegraphics{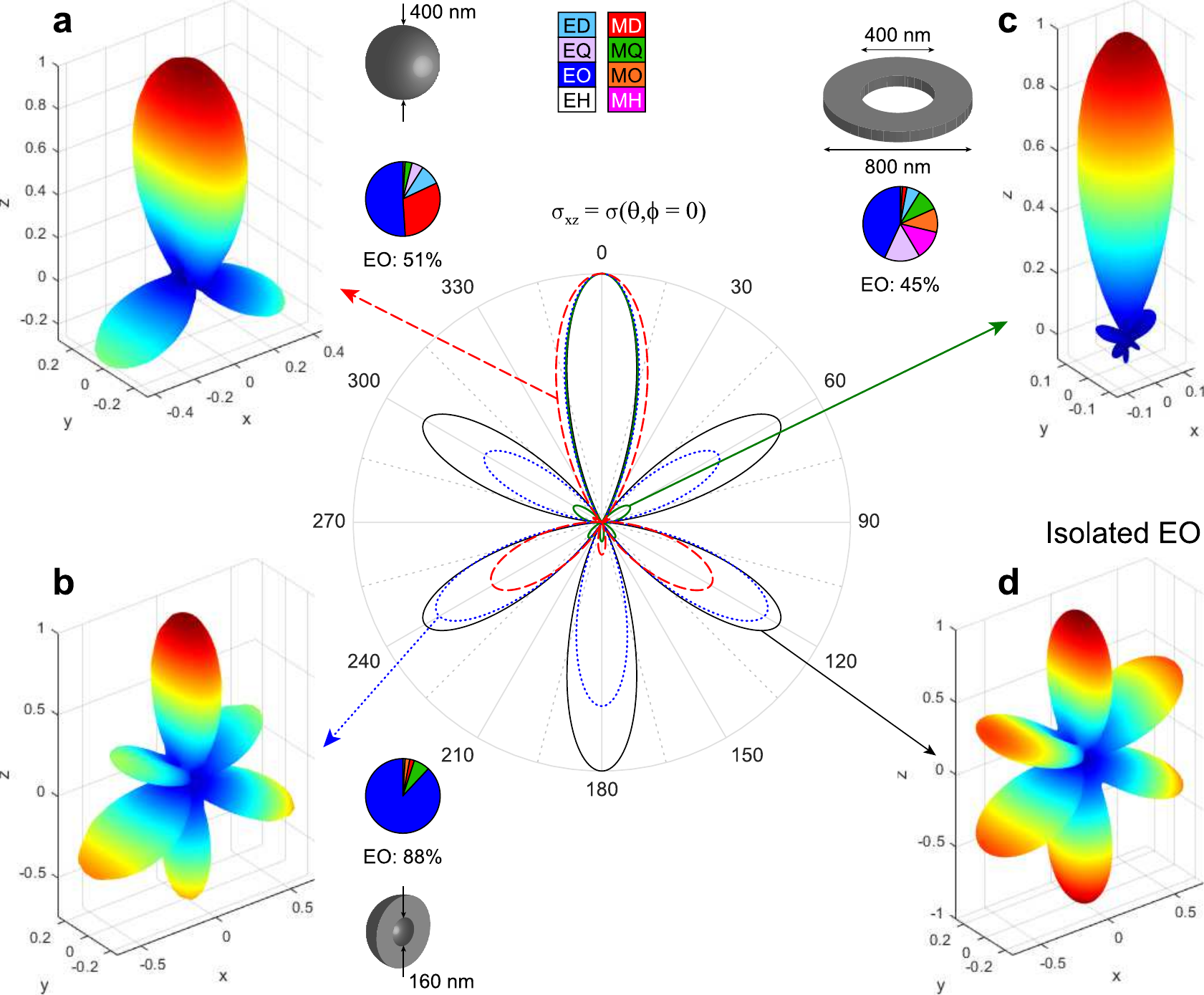}
  \caption{
	Deformation of the perfect electric octopole (EO) scattering diagram due to the interference with other multipoles contributions. (a-d) Numerically calculated scattering diagrams for a solid Si sphere at $\lambda$ = 810 nm (a), Si shell with a void diameter of 160 nm at $\lambda$ = 800 nm (b), Si ring with 400/800 nm inner/outer diameter at $\lambda$ = 800 nm (c), and compared with the radiation diagram of isolated electric octopole (d). Center: scattering polar plot at $\phi$ = 0 (i.e., in $xz$-plane). Corresponding spectra for chosen particles can be found in Figures~\ref{fig1}a,b and Figure~\ref{fig2}c. The outer diameter of Si sphere/shell is 400 nm, and the ring thickness is 80 nm. The refractive index of the surroundings and inside the void is $n$ = 1.48. In case of the solid sphere (a), the deconstructive interference between EO with MD leads to significant suppression of three out of the six major scattering lobes of EO. For this case $\sigma_{\rm{EO}} \approx 51\%$ and $\sigma_{\rm{MD}} \approx 31\%$, therefore the maximum expected asymmetry in forward/backward scattering is $\sim$9. In case of the ring (c), interference of EO with several other multipoles leads to a significant enhancement of the forward scattering lobe, thus resulting in a strong asymmetry in forward/backward scattering. Nevertheless, the angular position of all six scattering lobes in all cases (a-c) agrees well with the one of the isolated EO, which is an indication of the dominating EO contribution. 
	}
  \label{S9}
\end{figure}

\end{document}